\documentclass[aoas,preprint]{imsart}

\RequirePackage[OT1]{fontenc}
\RequirePackage[numbers]{natbib}

\usepackage{amsthm,amsmath}
\usepackage{amsbsy}

\input xy
\xyoption{all}

\usepackage{algorithm}
\usepackage[noend]{algpseudocode}

\usepackage{epsf}
\usepackage{epsfig}

\usepackage[labelfont=bf,labelsep=period,justification=justified]{caption}

\makeatletter
\renewcommand{\@biblabel}[1]{\quad#1.}
\makeatother

\date{}

\newcommand{\bfD}{\boldsymbol{D}}
\newcommand{\bfx}{\boldsymbol{x}}

\newcommand{\ind}{1\!\!1}
\newcommand{\bfy}{\boldsymbol{y}}
\newcommand{\bfX}{\boldsymbol{X}}

\newcommand{\bfl}{\boldsymbol{l}}
\newcommand{\bfw}{\boldsymbol{w}}
\floatstyle{plain}
\newfloat{supplefig}{htbp}{lgr}
\floatname{supplefig}{Figure S}

\begin{document}

\begin{frontmatter}

% "Title of the Paper"
\title{Reducing overfitting in challenge-based competitions}
\runtitle{Reducing overfitting in challenge-based competitions}

\author{Elias Chaibub Neto$^{1,\ast}$, Bruce R Hoff$^{1}$, Chris Bare$^{1}$, Brian M Bot$^{1}$, Thomas Yu$^{1}$, Lara Magravite$^{1}$, Andrew D Trister$^{1}$, Thea Norman$^{1}$, Pablo Meyer$^{2,\star}$, Julio Saez-Rodrigues$^{3,4,\star}$, James C Costello$^{5,\star}$, Justin Guinney$^{1,\star}$, Gustavo Stolovitzky$^{6, \star}$}

\footnote[0]{$^\ast$ Corresponding author. $^{1}$ Sage Bionetworks, Seattle, Washington, USA. $^{2}$ IBM Thomas J. Watson Research Center, Yorktown Heights, New York, USA. $^{3}$ European Molecular Biology Laboratory, European Bioinformatics Institute (EMBL-EBI), Wellcome Trust Genome Campus, Hinxton, UK. $^{4}$ RWTH-Aachen University Hospital, Joint Research Centre for Computational Biomedicine (JRC-COMBINE), Aachen, Germany. $^{5}$ Department of Pharmacology, University of Colorado Anschutz Medical Campus, Aurora, Colorado, USA. $^{6}$ IBM Translational Systems Biology and Nanobiotechnology, Yorktown Heights, New York, USA. $^\star$ These authors contributed equally to this work.}

\runauthor{Chaibub Neto et al.}

\begin{abstract}
Over-fitting is a dreaded foe in challenge-based competitions. Because participants rely on public leaderboards to evaluate and refine their models, there is always the danger they might over-fit to the holdout data supporting the leaderboard. The recently published Ladder algorithm aims to address this problem by preventing the participants from exploiting willingly or inadvertently minor fluctuations in public leaderboard scores during model refinement. In this paper, we report a vulnerability of the Ladder that induces severe over-fitting of the leaderboard when the sample size is small. To circumvent this attack, we propose a variation of the Ladder that releases a bootstrapped estimate of the public leaderboard score instead of providing participants with a direct measure of performance. We also extend the scope of the Ladder to arbitrary performance metrics by relying on a more broadly applicable testing procedure based on the Bayesian bootstrap. Our method makes it possible to use a leaderboard, with the technical and social advantages that it provides, even in cases where data is scant.
\end{abstract}

\end{frontmatter}

\section*{Introduction}

Challenge-based competitions\cite{bender2016,costello2013,boutros2014} have became a popular strategy for crowdsourcing and benchmarking solutions in complex data science problems. In these competitions, the community is presented with a challenge, the data to address the challenge, and a mechanism for independent and unbiased assessment of the submitted solutions. In biomedical research, challenge-based competitions have been used to address methodological bottlenecks in several areas including text mining\cite{trec2015}, gene network inference\cite{marbach2012}, prediction of drug synergy\cite{costello2014}, and prediction of protein structure, function, and interactions\cite{radivojac2013,moult2014}. The Dialogue for Reverse Engineering Assessments and Methods (DREAM) consortium, alone, has organized over 35 challenges examining complex prediction problems in biology and medicine (http://dreamchallenges.org). More generally, challenge organizing companies such as InnoCentive, have been crowdsourcing solutions for problems in engineering, computer science, math, chemistry, life sciences, physical sciences and business, while Kaggle specializes in predictive modeling and analytics competitions tailored, in most part, to industry problems, and Topcoder specializes in computer programming contests.

In predictive-modelling competitions, a participant is usually provided with data and is required to predict an output variable. The challenge organizers usually split all the available data into training, public leaderboard, and final scoring sets. Challenge participants have access to both input and response data in the training set, but only to the input variables in the public and final leaderboard sets. In the first phase of the challenge, participants generate and submit predictions to the public leaderboard, then the challenge organizers score and rank the participants according to their predictive performance. Multiple submissions per participant are usually allowed during this phase. The ranks of the participants in the public leaderboard are, nonetheless, not used in the determination of the challenge best performers. Rather, the purpose of public leaderboard is to provide the participants the opportunity to evaluate and refine their own models and compare their performance with others, prior to submitting their final predictions.

Because participants rely on sequential submissions to evaluate and refine their models, there is always the danger they might over-fit their models to the holdout response data supporting the public leaderboard. There may also be malicious attempts to deliberately over-fit the public leaderboard in order to discourage other participants from making a final submission, or to reverse engineer the response data in the public leaderboard to increase the amount of training data used to make the final submission. In either case, many challenges are one-of-a kind datasets that cannot be easily reproduced, such as challenges based on clinical trials or large cohorts of patients. In these cases, it is important to get the methodological aspects of challenge organization and rigor in scoring correctly, so that valuable datasets are used in the most advantageous ways to advance the field. To address these problems, participants are often limited in the number of submissions they can make to the leaderboard and performance scores are rounded to a less precise value. This practice is, nonetheless, far from ideal since it also restricts the ability of the participants to refine their models and generate better solutions to the problem the challenge is aiming to solve. In order to relax this constraint, Hardt and Blum\cite{ladder2015} recently proposed the Ladder algorithm (Algorithm \ref{alg:ladder} in Methods), which reduces over-fitting by preventing participants from exploiting minor fluctuations in public leaderboard scores during their model refinement activities (that is, preventing improvements to the hundreth or thousandth decimal place of the performance metric). Mechanistically, the Ladder only releases the actual (rounded) score of a new submission if the score presents a \textit{statistically significant} improvement over the previously best submission of the participant. If not, the Ladder releases the score of the participant's best submission to that point.

The Ladder mechanism is simple and represents a real improvement over the so called Full-disclosure mechanism (i.e., simply releasing the public leaderboard score for every submission made), which cannot protect less careful participants from over-fitting the public leaderboard response data. Hardt and Blum illustrate this point, in the context of a classification problem, showing how a variant of the Ladder (called parameter-free Ladder, and described in Algorithm \ref{alg:parfreeladder} in Methods) can withstand an aggressive boosting attack, to which the Full-disclosure mechanism is completely defenseless. In this paper, we focus on regression problems and evaluate the Ladder mechanism (Algorithm \ref{alg:ladderttest} in Methods) under two adversarial attacks. Both are inspired by Freedman's paradox\cite{freedman1983}, where the selection of the features entering a multiple regression model is guided by the public leaderboard. The first, denoted ``Freedman's attack", describes an efficient recipe to over-fit the public leaderboard response data. Our experiments show the effectiveness of the Ladder algorithm in this context. The second attack, on the other hand, is based on a more aggressive step-forward variation of Freedman's attack (denoted the ``step-forward Freedman's attack"), and can lead to severe over-fitting of the Ladder leaderboard when the sample size is small. The step-forward attack exploits a vulnerability of the Ladder, namely, that it leaks too much information regarding the holdout data when it releases the public leaderboard score of the best model to that point. To circumvent this problem, we propose an important variation of the Ladder, called the LadderBoot algorithm, which releases a bootstrapped estimate of the public leaderboard score, instead of the actual (rounded) score. In our experiments, the LadderBoot compared favorably to the Ladder. A detailed description of the LadderBoot mechanism is given in Algorithm \ref{alg:ladderbootttest} in Methods.

An additional contribution of this paper is to extend the scope of the Ladder and LadderBoot mechanisms to performance metrics other than ones corresponding to empirical risk estimators, to which the Ladder algorithm is not directly applicable. Our more general algorithms rely on a more broadly applicable testing procedure based on the Bayesian bootstrap and we call the BayesBootLadder and BayesBootLadderBoot mechanisms (Algorithms \ref{alg:bayesbootladder} and \ref{alg:bayesbootladderboot} in Methods).

\section*{Results}

We experimented with variations of the Ladder mechanism using a subset of the data from the Alzheimer Disease Big Data DREAM Challenge 1\cite{allen2016}, denoted as ``AD Challenge". The goal of this challenge was to predict mini-mental state examination (MMSE) scores\cite{mmse1975} using 2,150 features (morphometric measurements) extracted from MRI data, as well as, clinical and demographic covariates. Here, we focused on the imaging features of $n = 628$ samples from the Alzheimer's Disease Neuroimaging Initiative (ADNI) data\cite{adni2015}. Also, for the sake of computational speed, we restricted most of our analyses to $p = 300$ randomly chosen imaging features. For some of the illustrative examples we also relied on a simulated data set with $n = 120$ and $p = 1,000$. To mirror real word examples, we employed a small sample size in these simulated examples that clearly show over-fitting and provide clear-cut illustrations of the issues.

In all experiments we split the data into training, public leaderboard, and final scoring sets, of roughly equal sample sizes. For each set we scaled both the features and response data. In order to assess the variability of the results, the experiments were replicated a large number of times (100 or 1000 times). Experiments were performed in two distinct ways: ``permuted response" experiments, where we kept the training, public leaderboard, and final/test sets fixed, but permuted the response data on each of these sets; and ``random data splitting" experiments, where the data was randomly split into training, public leaderboard, and final scoring set.

Note that any signal detected in the ``permuted data" experiments is due purely to random chance, and amounts to over-fitting the model to noise fluctuations in the holdout set. Hence, the permuted responses experiments provide the clearest illustrations of over-fitting. Results from the random data split experiments, on the other hand, are not as clear cut since the detected signal might be due to real associations, to over-fitting, or (most likely) to a combination of both. In any case, we investigated (and reported in the Supplement) the performance of the algorithms in the random data split experiments, as these experiments reproduce the real conditions under which these methods are usually applied.

\subsection*{Freedman's attack}

This attack is inspired by ``Freedman's paradox", which portrays a problem in model selection where features not related to the response variable can appear artificially important\cite{freedman1983}, and describes a recipe to over-fit the public leaderboard in the context of a regression problem. The idea is to perform feature selection guided by the scores released to the public leaderboard. Explicitly, the attack proceeds as follows. First, we fit separate univariate regression models of the response on each of the $p$ features using the training data, then submit all $p$ univariate models to the public leaderboard. Next, we rank the univariate models according to the public leaderboard scores, and train a multiple linear regression model using the top $k$ features (for an arbitrary choice of $k$). We then evaluate and compare the performances of this final multiple regression model on the public leaderboard and final scoring set.

\begin{figure}[!h]
\begin{center}
\includegraphics[angle=270, scale = 0.49, clip]{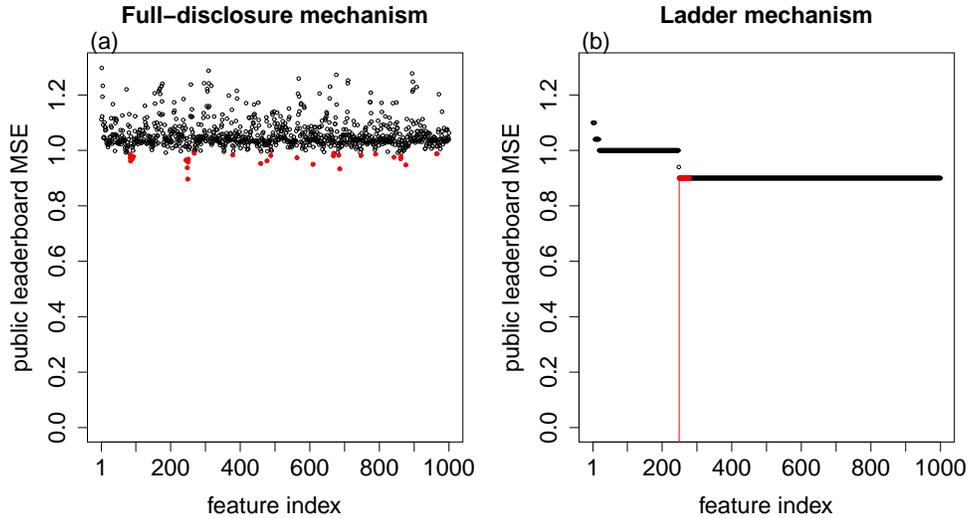}
\caption{Public leaderboard MSE scores released by the Full-disclosure mechanism (panel a) and the Ladder mechanism based on $\alpha = 0.15$ (panel b). Results based on a simulated data set with shuffled response data. The red dots represent the 30 features with smallest MSE scores. The vertical red line in panel b shows the index of the feature responsible for the last ``jump in score" in the public scores released by the Ladder mechanism (feature number 249 in this example).}
\label{fig:standardvsladderreleasedscores}
\end{center}
\end{figure}

Fig.\ref{fig:standardvsladderreleasedscores} presents the public leaderboard mean squared error (MSE) released by the Full-disclosure mechanism (Fig.\ref{fig:standardvsladderreleasedscores}a) and the Ladder mechanism (Fig.\ref{fig:standardvsladderreleasedscores}b) for a simulated data set with $p = 1000$ features, $n = 120$ samples, and significance level associated with the one sided t-test in Algorithm \ref{alg:ladderttest} set to $\alpha = 0.15$. We permuted the response data so that none of the features are truly associated with the response. The features were simulated from a multivariate normal distribution with strong correlation structure. Since we scaled the response data, models with an MSE greater than 1 are performing worse than the intercept only model (where the prediction in the leaderboard set is given by the mean response value in the training set), whereas models with MSE smaller than 1 are performing better. From Fig.\ref{fig:standardvsladderreleasedscores}a we see that most of the features generate MSE slightly above 1 (with a few considerably larger or smaller than 1 due to random chance). The red dots represent the $k = 30$ features with smallest MSE scores. Note that while the red dots are spread across the 1,000 features in Fig.\ref{fig:standardvsladderreleasedscores}a, they are all clustered together on Fig.\ref{fig:standardvsladderreleasedscores}b (corresponding to feature 249 and the following 29 features), since the scores released by the Ladder are identical for all features after index 249 (by convention, we simply select the first $k$ consecutive features, instead of randomly selecting the top features among all tied features). It is clear from Fig.\ref{fig:standardvsladderreleasedscores} that the multiple regression model including the top features from the Full-disclosure public leaderboard includes 30 features with some degree of association to the permuted response data, whereas the features selected from the Ladder leaderboard contain only a few features associated with the permuted responses (including feature 249 and, possibly, a few neighboring features due to the strong correlation of the simulated features) and won't be able to drastically over-fit the public leaderboard.

\begin{figure}[!h]
\begin{center}
\includegraphics[angle=270, scale = 0.78, clip]{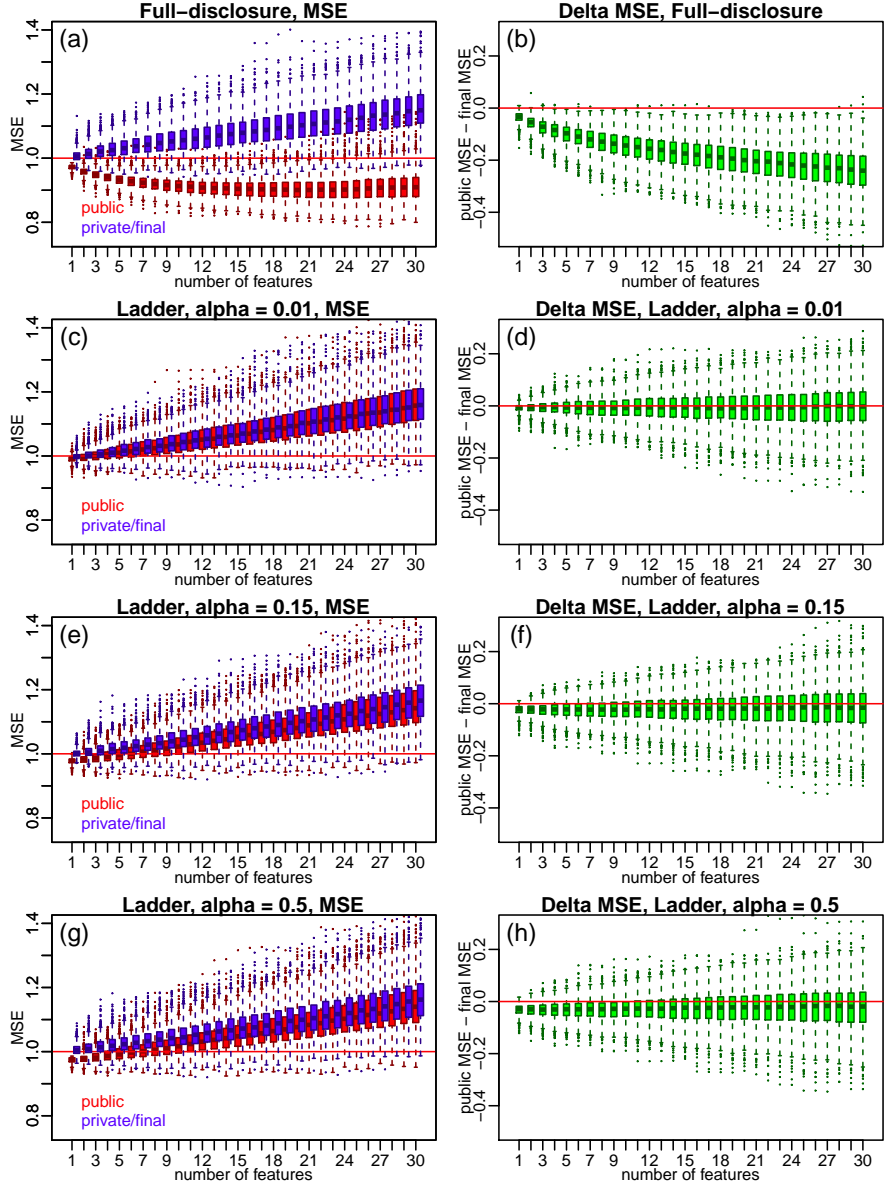}
\caption{Comparison of the Full-disclosure vs Ladder mechanisms (based on $\alpha$ equal to 0.01, 0.15, and 0.5) using the AD Challenge data. Results based on 1,000 response data permutations, and on a subset of 300 randomly chosen features.}
\label{fig:standarvsladderpermuted}
\end{center}
\end{figure}

Fig.\ref{fig:standarvsladderpermuted} presents the results of the permuted response experiment comparing the Full-disclosure and Ladder mechanisms (for varying $\alpha$ levels) under Freedman's attack. The boxplots in the left column panels represent the distributions of the MSEs released by the public (red) and final (blue) leaderboards, while right column panels present the boxplots of the $\Delta \mbox{MSE}$,
$$
\Delta \mbox{MSE} = \mbox{MSE}_{public} - \mbox{MSE}_{final}~,
$$
generated from 1,000 random permutations of the response data, for $k$ increasing from 1 to 30. A negative $\Delta$MSE indicates over-fitting to the public leaderboard data. Fig.\ref{fig:standarvsladderpermuted}a and b show the inability of the Full-disclosure mechanism to withstand Freedman's attack, with MSEs in the public leaderboard decreasing as a function of $k$, while the final MSEs actually increase with $k$. Fig.\ref{fig:standarvsladderpermuted}c and d, on the other hand, clearly show the effectiveness of the Ladder mechanism for avoiding over-fitting ($\alpha = 0.01$, red boxplots closely track the blue boxplots). Fig.\ref{fig:standarvsladderpermuted}e to h, illustrate how the Ladder can still withstand Freedman's attack when employed with more relaxed significance levels. Note that even though we observe an increasing amount of over-fitting for increasing $\alpha$ levels, the amount is nevertheless small, and the results seem to be robust to the choice of $\alpha$. We observe similar, although not as extreme, results for the random splitting experiments (Fig.S\ref{fig:standardvsladderoriginal}).

\subsection*{Step-forward Freedman's attack}

Even though the Ladder mechanism seems to be effective against Freedman's attack, it is nonetheless defenseless against the more aggressive attack we describe next. This second attack assumes that the challenge organizers do not restrict the number of allowed submissions per participant. It is based on a variation of Freedman's attack and exposes a vulnerability of the Ladder mechanism. The problem is that whenever a model is better (by a given margin) than the best model to this point, the Ladder mechanism releases the actual (rounded) public leaderboard score of the model. By doing so, it leaks the information that this particular feature has some predictive power. As illustrated in Fig.\ref{fig:standardvsladderreleasedscores}b, an attacker can easily select a good (although not necessarily the best) univariate feature for predicting the public leaderboard response data, by simply tracking out the feature responsible for the last ``jump" in performance as reflected in the leaderboard and released by the Ladder mechanism. Once the first feature is selected, the attacker can engage in an iterative attack, mimicking the step-forward best subset feature selection approach. In a second iteration the attacker fits $p - 1$ separate regression models (each including the feature selected in the first step plus one of the $p - 1$ remaining features) and then selects a new feature responsible for yet another last jump in score in Ladder's public leaderboard. In the third iteration, the attacker fits $p - 2$ separate regression models (each including the two features selected in the two previous steps, plus one of the $p - 2$ remaining features) and so on. Algorithm \ref{alg:stepforwardattack} in Methods describes the step-forward attack in more detail.

\begin{figure}[!h]
\begin{center}
\includegraphics[angle=270, scale = 0.425, clip]{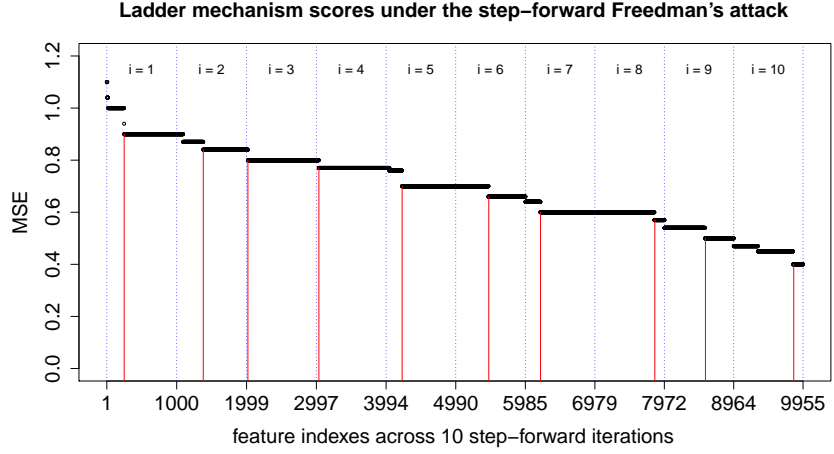}
\caption{Public leaderboard scores released by the Ladder mechanism ($\alpha = 0.15$) under a step-forward attack run for 10 iterations. The red vertical lines show the positions of the features responsible for the last jump in score in each one of the 10 iterations. Results based on a simulated data set.}
\label{fig:ladderreleased10steps}
\end{center}
\end{figure}
\begin{figure}[!h]
\begin{center}
\includegraphics[angle=270, scale = 0.425, clip]{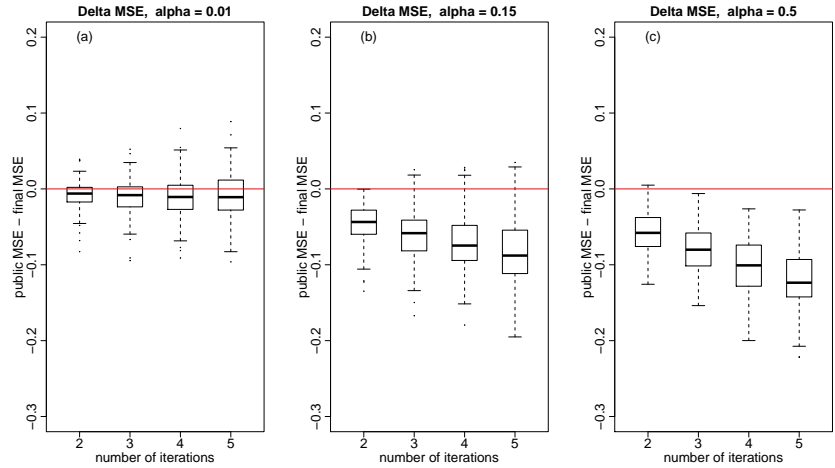}
\caption{Over-fitting incurred by the Ladder mechanism (based on $\alpha$ equal to 0.01, 0.15, and 0.5) under step-forward attacks with increasing number of iterations, using the AD Challenge data. Results based on 100 permutations of the response data, and on a subset of 300 randomly chosen features.}
\label{fig:stepforwardladderpermuted}
\end{center}
\end{figure}

Fig.\ref{fig:ladderreleased10steps} illustrates a step-forward attack (ran for 10 iteration steps) to the same simulated data-set employed in the generation of Fig.\ref{fig:standardvsladderreleasedscores}. After 10 iterations, we obtain a model (containing 10 features) with a MSE around 0.4 in the Ladder's public leaderboard, whereas the score on the final scoring set is expected to be close to 1 (none of the features in this simulated data-set are truly associated with the response). Contrary to Freedman's attack, where the final multiple regression model derived from the Ladder leaderboard contained the feature responsible for the last jump in addition to a number of features that were not actually associated with the permuted response, all 10 features (red vertical lines in Fig.\ref{fig:ladderreleased10steps}) selected during the step-forward attack are associated (by pure chance) to the permuted response data, so that the respective multiple regression model can strongly over-fit the public leaderboard holdout data.

In order to further illustrate the amount of over-fitting incurred by the Ladder mechanism under the step-forward attack, we performed a permuted responses experiment using the AD challenge data. Fig.\ref{fig:stepforwardladderpermuted} presents the boxplots of the $\Delta$MSE for step-forward attacks based on 2, 3, 4, and 5 iteration steps under three different $\alpha$ levels (0.01, 0.15, and 0.50). We observe similar results for the random splitting experiments (Fig.S\ref{fig:stepforwardladderoriginal}).

As expected, we observe an increase in the amount of over-fitting as we increase the number of iterations of the step-forward algorithm. Furthermore, as clearly shown by the considerable smaller amount of over-fitting incurred by smaller $\alpha$ values (Fig.\ref{fig:stepforwardladderpermuted}a-c, the results from the step-forward attack seem to be more sensible to the choice of $\alpha$ than the results from Freedman's attack. This increased sensibility is, nonetheless, not surprising since smaller $\alpha$ values lead to a more conservative mechanism with fewer but larger jumps, whereas larger $\alpha$ values lead to more liberal mechanisms with frequent but smaller jumps and can increase the chance of over-fitting during an iterative attack.

\subsection*{The LadderBoot mechanism}

%Fortunately, there is a way to mitigate the information leakage incurred by the Ladder mechanism. A natural tweak of the Ladder mechanism capable to mitigate the information leakage issue in the Ladder algorithm is to simply add some randomness to the output of the Ladder mechanism. By doing so, the attacker is no longer able to easily pinpoint features with minor incremental value. To this end, we adopt a bootstrap based privacy mechanism, where instead of releasing the public leaderboard score, we release the average over $b$ bootstrap replications of the loss vectors data. A detailed description of our algorithm, denoted the LadderBoot mechanism, is provided on Algorithm \ref{alg:ladderbootttest} in Methods.

In order to mitigate the information leakage incurred by the Ladder mechanism, we add some randomness to the output of the algorithm, so that the attacker is no longer able to easily pinpoint features with minor incremental value. To this end, we adopt a bootstrap based privacy mechanism, where instead of releasing the public leaderboard score, we release the average over $b$ bootstrap replications of the loss vectors data. By doing so, the algorithm effectively employs a more robust measure of performance, less dependent on the strict composition of the leaderboard test set, and able to obscure the information that can be used for overfitting. A detailed description of our algorithm, denoted the LadderBoot mechanism, is provided on Algorithm \ref{alg:ladderbootttest} in Methods.

%By doing so, the attacker is no longer able to easily pinpoint features with minor incremental value. To this end, we adopt a bootstrap based privacy mechanism, where instead of releasing the public leaderboard score, we release the average over $b$ bootstrap replications of the loss vectors data. A detailed description of our algorithm, denoted the LadderBoot mechanism, is provided on Algorithm \ref{alg:ladderbootttest} in Methods.

An important practical question is to determine the amount of randomness that would give the attacker a hard time, but at the same time, would allow the algorithm to release a useful score, capable of guiding challenge participants on their model building process. Large values of $b$ lead to precise estimates of the score but can also make it easy to detect features with some predictive power. Small $b$ values, on the other hand, lead to less precise estimates, but make it harder to detect informative features. In addition to the number of bootstraps, the adopted $\alpha$ value controls the average gap between jumps, since the larger the gap, the greater the amount of randomness needed to make the step-forward attack difficult.

\begin{figure}[!h]
\begin{center}
\includegraphics[angle=270, scale = 0.47, clip]{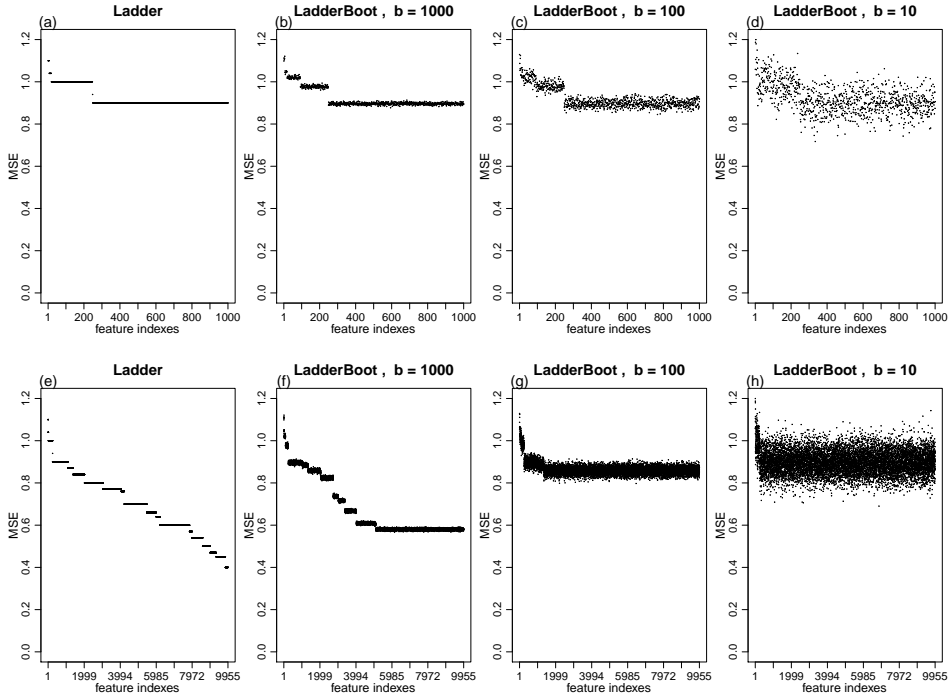}
\caption{Comparison of the public leaderboard scores released by the Ladder and LadderBoot mechanisms (for $b$ equal to 1,000, 100, and 10, and $\alpha = 0.15$) under Freedman's attack (panels a to d) and under the step-forward attack ran for 10 iterations (panels e to h). Results based on simulated data.}
\label{fig:ladderladderboot1and10steps}
\end{center}
\end{figure}

Fig.\ref{fig:ladderladderboot1and10steps}a-d compares the released public leaderboard scores of the Ladder algorithm against the LadderBoot mechanism, under Freedman's attack, using $b$ equal to 1000, 100, and 10. In all four cases, $\alpha = 0.15$. We see that the amount of bootstrap noise generated with 1000 (or even 100) bootstrap replications is not enough to hide the last performance jump position. On the other hand, adoption of $b = 10$ makes it difficult to determine the last jump location. Fig.\ref{fig:ladderladderboot1and10steps}e-h, show the released public leaderboard scores generated by the step-forward attack ran for 10 iterations. Clearly, the amount of over-fitting decreases as we decrease the number of bootstrap replications. At each iteration step, the determination of the last performance gain location was done automatically using the binary segmentation\cite{binseg1974} method implemented in the \texttt{changepoint}\cite{changepoint2014} R package\cite{r2014}. The reduction in over-fitting (i.e., the plateauing at higher MSEs) shows that the bootstrap noise prevented the change-point method from detecting the correct position of the jumps, even though we informed the true number of jumps to the change-point detection method at each iteration of the step-forward attack.

\begin{figure}[!h]
\begin{center}
\includegraphics[angle=270, scale = 0.59, clip]{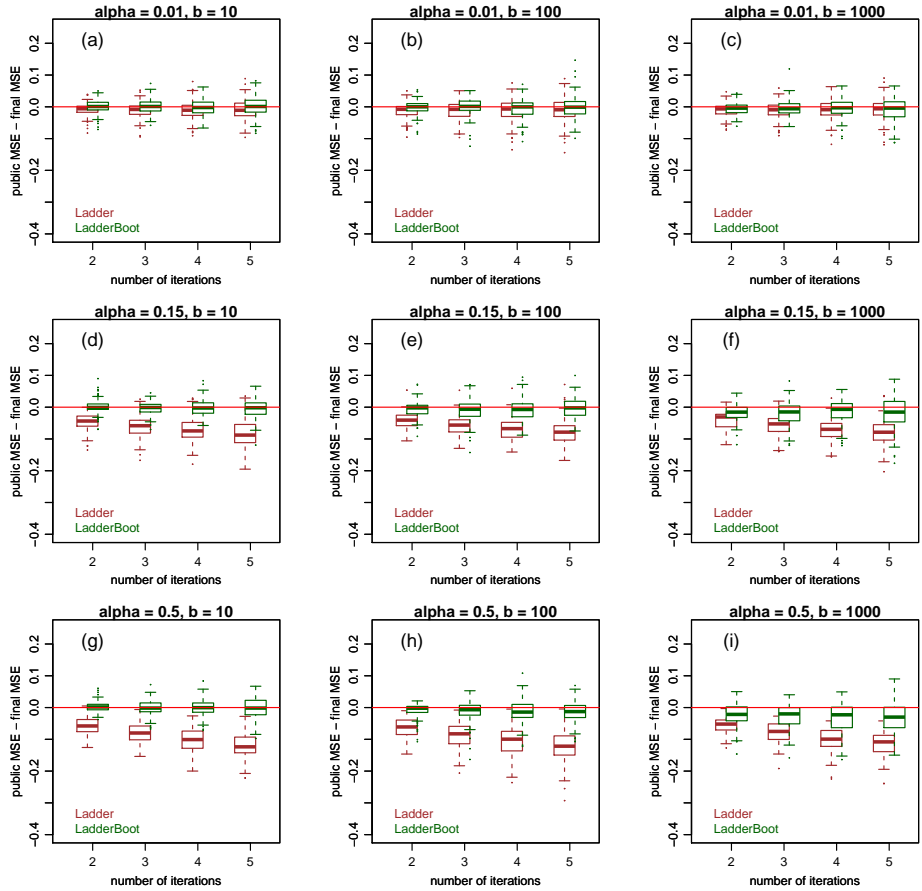}
\caption{Comparison of the $\Delta$MSE values released by the Ladder and LadderBoot algorithms, for all combinations of $\alpha = \{ 0.01, 0.15, 0.5 \}$ and $b = \{ 10, 100, 1000\}$, using the AD Challenge data. Results based on 100 permutations of the response data, and on 300 randomly selected features.}
\label{fig:fig8}
\end{center}
\end{figure}

Fig.\ref{fig:fig8} compares the $\Delta$MSE of the Ladder and LadderBoot mechanisms for all combinations of $\alpha = \{ 0.01, 0.15, 0.5 \}$ and $b = \{ 10, 100, 1000\}$ in a permuted responses experiment. The results show that the LadderBoot mechanism outperforms the Ladder mechanism under the step-forward attack in all cases investigated. It also shows that for larger values of $b$, the LadderBoot mechanism can also over-fit the public leaderboard data (especially for large values of $\alpha$). We observe similar results for the random splitting experiments (Fig.S\ref{fig:fig9}). Fig.S\ref{fig:fig10} and S\ref{fig:fig11} show the public leaderboard MSE scores for the permuted responses and random data split experiments, respectively. For completeness, we present a comparison of the Ladder and LadderBoot mechanisms under Freedman's attack in Suppl. Text S1.

\subsection*{Extension to additional scoring metrics}

The decision to release a new performance measure in the Ladder and LadderBoot algorithms is implicitly guided by a one-sided paired t-test of loss vectors (equation \ref{eq:pairedttest} in Methods). The adoption of this implicit test restricts the applicability of the Ladder and LadderBoot mechanisms to performance metrics corresponding to empirical risk estimators such as mean squared error and mean absolute error, but cannot be directly applied to commonly used metrics such as Lin's concordance correlation coefficient\cite{lin1989} and Pearson's correlation, since these metrics cannot be directly expressed as an arithmetic average of element-wise loss values.

In order to extend the applicability of the Ladder and LadderBoot mechanisms to general scoring metrics, we replace the paired t-test by the Bayesian bootstrap\cite{rubin1981}. We denote these Bayesian bootstrap based variations of the Ladder and LadderBoot algorithms as the BayesBootLadder and BayesBootLadderBoot mechanisms, respectively. Algorithms \ref{alg:bayesbootladder} and \ref{alg:bayesbootladderboot} in the Methods, present in detail these mechanisms. An illustration based on Lin's concordance correlation coefficient\cite{lin1989} is presented in the Supplementary Text S2.

\section*{Discussion}

In this paper we show how the Ladder algorithm can leak too much information about the public leaderboard performances and is vulnerable to a step-forward variation of Freedman's attack when the sample size is small. We also propose the LadderBoot mechanism and illustrate how the adoption of a bootstrap privacy mechanism is able to mitigate this vulnerability by adding a controlled amount of noise to the public leaderboard. Bootstrap based privacy mechanisms have been previously used in different contexts, including in the selection of micro-data records to be disclosed to the public\cite{fienberg1994,fienberg1996,raghunathan2003, melville2012}, masking of contingency tables\cite{heer1993}, disclosure limitation in regression analysis\cite{jones1998}, and response masking in remote analysis systems\cite{muralidhar2015}. We point out, nonetheless, that the combination of bootstrapping and Ladder into a single mechanism is novel.

We also extended the applicability of the Ladder and LadderBoot mechanisms to more general scoring metrics (other than the ones corresponding to empirical risk estimators) by replacing the one-sided paired t-test employed in these algorithms by a Bayesian bootstrap based test. It is not only possible, but actually advisable to use the Bayesian bootstrap test for metrics corresponding to empirical risk estimators since the bootstrap approach does not require any of the distributional assumptions made by the paired t-test (it only requires independence of the data samples). The price payed for this more general applicability is, nonetheless, the increased amount of computation.

One possible refinement of the algorithms proposed in this paper is to consider adaptive variations, where the initial submissions of each participant is perturbed by a small amount of noise that is adaptively increased as a function of the number of submissions provided by the participant. For example, starting with a large $b$ and decreasing $b$ as the number of submissions increases. Such adaptive algorithms would be able to neutralize potentially less careful participants while minimally penalizing the model refinement activities of more parsimonious participants.

It is important to highlight that even though the LadderBoot and BayesBootLadderBoot mechanisms are able to better withstand a step-forward attack, in comparison to the Ladder, they are still vulnerable to over-fitting if we do not restrict the number of allowed submissions per participant. For instance, even if we adopt a LadderBoot (BayesBootLadderBoot) mechanism based on a single bootstrap replication ($b = 1$), an attacker can still figure out the last ``performance jump" positions if allowed an unlimited number of submissions to the public leaderboard. All an attacker has to do is submit each univariate model a large number of times and compute the average of the LadderBoot (BayesBootLadderBoot) released scores. More concretely, the attacker could (starting with the first feature): (1) generate the feature's univariate model; (2) submit the univariate model to the public leaderboard a large number of times, $T$; (3) compute and record the average of the released LadderBoot (BayesBootLadderBoot) scores; (4) repeat steps 1 to 3 for each one of the remaining features; and (5) plot the recorded averages against the feature's indexes. For a large $T$, the average scores will likely be very close to the public leaderboard score released by the Ladder (BayesBootLadder), and the attacker should be able to easily determine the ``last jump" position, and proceed with a respective ``average across repeated submissions" step-forward attack. A simple solution to this issue would be to check the submissions to see if they are the same and, if so, reject the submission.

These considerations show that the LadderBoot mechanism, although more protective than the Ladder, still leaks some information about the public leaderboard scores (information leakage is really unavoidable if the released score is to be useful). In any case, the Ladder, LadderBoot and BayesBootLadderBoot algorithms provide a reliable mechanism for maintaining an accurate public leaderboard, as long as we impose a restriction on the number of allowed submissions per participant, or monitor that the submissions are not identical or tightly correlated. The question then becomes: how many submissions can we safely allow, before participants start over-fitting the public leaderboard? Recent theoretical work by Dwork et al\cite{dwork2015} answered this exact question, but for a distinct differential privacy algorithm called Thresholdout. In the original Ladder paper, Hardt and Blum\cite{ladder2015} derive a theoretical worst-case upper bound for the leaderboard accuracy of a bounded metric in the interval $[0, 1]$, in the context of a classification problem. Their results are not directly applicable to the regression problems based on unbounded metrics considered in the present work, or to the LadderBoot and BayesBootLadderBoot algorithms. Hence, the determination of the maximum number of submissions to the public leaderboard remains an open theoretical question in our context.

It might still be possible to get some empirical guidance on this matter, by experimenting with tuning parameter choices during the dry-run phase of a challenge, where organizers curate the data, develop and refine the questions to be possed to the community, and assess the feasibility of the challenge. In Supplementary Text S3 we describe an heuristic approach for determining a general ballpark for the number of submissions. It is important, nonetheless, to indicate that these empirical results can only provide a rough guideline, as the amount of over-fitting might also depend on the scoring metric and on the models employed during the over-fitting quantification study. (In Supplementary Text S4 we illustrate how the amount of variability of the released scores also depends on the quality of predictions themselves, so that results might be different depending on whether we employ a simple multiple linear regression model, or use elastic-net or random forests, when combining the top features selected during the attacks.)

In any case, we argue that even rough empirical estimates can be valuable in practice, since we might be able to scale up the number of allowed submissions, even when we are very conservative and adopt a much smaller limit than our empirical results suggest to be safe. These empirical evaluations are particularly important for challenges employing small sample sizes, as these are particularly prone to over-fitting.  Finally, it is important to highlight that these estimates are based on very aggressive and malicious attacks, but in practice most challenge participants will not behave this way.

In summary, the adoption of the Ladder or its variations will lower the risk to scale up the number of allowed submissions to the public leaderboard. For instance, instead of allowing only 3 submissions, a challenge organizer might be able to safely allow, say, 30 or 300 submissions depending on whether he/she adopts the Ladder or the LadderBoot leaderboard. This represents an important improvement, as it provides participants the opportunity, and better conditions, to refine their models and ultimately generate better solutions to the problem the challenge is aiming to solve. The present work was motivated by the need to control over-fitting in biomedical research challenges, where the amount of data is sometimes scant. We point out, nonetheless, that the proposed methodology is general and applicable to crowdsourced benchmarking of predictive models in engineering, life sciences, physical sciences and business. We plan to implement these algorithms in future DREAM challenges.

\section*{Methods}

Algorithm \ref{alg:ladder} presents the general Ladder mechanism proposed by Hardt and Blum \cite{ladder2015}. Here, $\bfD = (\bfX, \bfy) = \{ (\bfx_1, y_1), \ldots, (\bfx_n, y_n) \}$ represents a data set of $n$ independent and identically distributed data points, $(\bfx_i, y_i)$, where $y_i$ corresponds to the output (response) variable, and $\bfx_i = (x_{i1}, \ldots, x_{ip})^T$ represents the vector of input (covariate) variables. The empirical risk of a prediction $\hat{\bfy} = f(\bfX)$ is defined as,
\begin{equation}
R(\hat{\bfy}) = \frac{1}{n} \sum_{i=1}^{n} l(f(\bfx_i), y_i)~,
\label{eq:empiricalrisk}
\end{equation}
where $l_i = l(f(\bfx_i), y_i)$ represents the loss associated with the prediction $\hat{y}_i$ of the outcome value $y_i$. We define a parameter $\eta$, which represents the margin (or step size) by which the current prediction needs to outperform the best prediction so far in order for the Ladder mechanism to release a new (rounded) score. Explicitly, if the condition on line 6 holds, the algorithm releases the rounded empirical risk value $[R(\hat{\bfy})]_{\eta}$, with the rounding precision controlled by $\eta$ as well (line 7). For instance, if the empirical risk is given by $R(\bfy_t) = 0.8763$, then the released scores are given by $[0.8763]_{0.1} = 0.9$, $[0.8763]_{0.01} = 0.88$, $[0.8763]_{0.001} = 0.876$, for $\eta$ equal to 0.1, 0.01, and 0.001, respectively. If the condition on line 6 does not hold, the algorithm releases the best score so far.

\begin{algorithm}
\caption{General Ladder mechanism (Hardt and Blum, 2015)}\label{alg:ladder}
\begin{algorithmic}[1]
\State \textbf{Input}: Data, $\bfD = \{ (\bfx_1, y_1), \ldots, (\bfx_n, y_n) \}$ \\
Assign initial estimate $R_0 \leftarrow \infty$.
\For{$t = 1, 2, \ldots$}
  \State Receive prediction $\hat{\bfy}_t$
  \State Compute empirical risk $R(\hat{\bfy}_t)$
  \If{$R(\hat{\bfy}_t) < R_{t-1} - \eta$}
    \State $R_t \leftarrow [R(\hat{\bfy}_t)]_{\eta}$
  \Else
    \State $R_t \leftarrow R_{t-1}$
  \EndIf
  \State Release $R_t$
\EndFor
\end{algorithmic}
\end{algorithm}

As pointed in \cite{ladder2015}, it is difficult to choose a fixed $\eta$ value that would work well throughout the challenge. To circumvent this problem, Hardt and Blum\cite{ladder2015} proposed the parameter-free Ladder algorithm (Algorithm \ref{alg:parfreeladder}), which adaptively finds a suitable margin value according to previous submissions to the algorithm by choosing $\eta = \mbox{sd}(\bfl_t - \bfl_{t-1})/\sqrt{n}$, where sd represents the standard deviation operator, and $\bfl_t = (l_{t,1}, \ldots, l_{t,n})^t$ and $\bfl_{t-1} = (l_{t-1,1}, \ldots, l_{t-1,n})^t$ represent, respectively, the loss vectors associated with the current and so far best predictions. Note that the condition on line 8 of Algorithm \ref{alg:parfreeladder} can be (approximately) re-expressed in the format of a paired t-test statistic,
$$
\frac{ \sqrt{n} \, (\bar{l}_t - [\bar{l}_{t-1}]_{1/n})}{\mbox{sd}(\bfl_t - \bfl_{t-1})} \, < \, -1
$$
where $\bar{l}_t = R(\hat{\bfy}_t) = n^{-1} \sum_{i=1}^{n} l_{t,i}$ and $\bar{l}_{t-1} = R(\hat{\bfy}_{t-1}) = n^{-1} \sum_{i=1}^{n} l_{t-1,i}$ represent the empirical risks. Clearly, the condition is (approximately) equivalent to performing a one-sided paired t-test adopting a p-value close to 0.15. (It would be exactly equivalent to a paired t-test if we had $\bar{l}_{t-1}$ in place of $[\bar{l}_{t-1}]_{1/n}$.) As pointed out by Hardt and Blum\cite{ladder2015}, this test employs increasingly smaller margin values as the best prediction (so far) gets increasingly accurate.

\begin{algorithm}
\caption{Parameter-free Ladder algorithm (Hardt and Blum, 2015)}\label{alg:parfreeladder}
\begin{algorithmic}[1]
\State \textbf{Input}: Data, $\bfD = \{ (\bfx_1, y_1), \ldots, (\bfx_n, y_n) \}$ \\
Assign initial estimate $R_0 \leftarrow \infty$ \\
Assign initial loss vector $l_0 = (0)_{i=1}^{n}$.
\For{$t = 1, 2, \ldots$}
  \State Receive prediction $\hat{\bfy}_t$
  \State Compute loss vector $\bfl_t \leftarrow (l(\hat{y}_{t,i} \, , \, y_i))_{i=1}^n$
  \State Compute $R(\hat{\bfy}_t) \leftarrow n^{-1} \sum_{i=1}^{n} l_{t,i}$
  %\State Compute $R_{t-1} \leftarrow n^{-1} \sum_{i=1}^{n} l_{t-1,i}$
  \If{$R(\hat{\bfy}_t) < R_{t-1} - \mbox{sd}(\bfl_t - \bfl_{t-1})/\sqrt{n}$}
    \State $R_t \leftarrow [R(\hat{\bfy}_t)]_{1/n}$
  \Else
    \State $R_t \leftarrow R_{t-1}$
    \State $\bfl_t \leftarrow \bfl_{t-1}$
  \EndIf
  \State Release $R_t$
\EndFor
\end{algorithmic}
\end{algorithm}

In this paper, nonetheless, we consider a slight extension of the parameter-free algorithm, where a challenge organizer can specify the significance level $\alpha$ for the underlying paired t-test (Algorithm \ref{alg:ladderttest}). Since this extended algorithm is no longer parameter free, we call it the Ladder algorithm, but point out that it is still a variant of the general Ladder mechanism (Algorithm \ref{alg:ladder}), and should not be confused with it. The main difference, relative to the parameter-free Ladder algorithm, is that the condition on line 9 can be re-expressed in the format of a one-sided paired t-test statistic,
\begin{equation}
\frac{ \sqrt{n} \, (\bar{l}_t - [\bar{l}_{t-1}]_{1/n})}{\mbox{sd}(\bfl_t - \bfl_{t-1})} \, < \, -c_\alpha
\label{eq:pairedttest}
\end{equation}
where $c_\alpha = F^{-1}_{n-1}(\alpha)$ represents the $\alpha$ quantile of a t-distribution with $n-1$ degrees of freedom.

\begin{algorithm}
\caption{Ladder algorithm (based on the paired t-test statistic)}\label{alg:ladderttest}
\begin{algorithmic}[1]
\State \textbf{Input}: Data, $\bfD = \{ (\bfx_1, y_1), \ldots, (\bfx_n, y_n) \}$ \\
Significance level $\alpha$ for underlying paired t-test, or corresponding critical value $c_\alpha = F^{-1}_{n-1}(\alpha)$ \\
Assign initial estimate $R_0 \leftarrow \infty$ \\
Assign initial loss vector $l_0 = (0)_{i=1}^{n}$.
\For{$t = 1, 2, \ldots, t_{max}$}
  \State Receive prediction $\hat{\bfy}_t$
  \State Compute loss vector $\bfl_t \leftarrow (l(\hat{y}_{t,i} \, , \, y_i))_{i=1}^n$
  \State Compute $R(\hat{\bfy}_t) \leftarrow n^{-1} \sum_{i=1}^{n} l_{t,i}$
  \If{$R(\hat{\bfy}_t) < R_{t-1} - c_\alpha \; \mbox{sd}(\bfl_t - \bfl_{t-1})/\sqrt{n}$}
    \State $R_t \leftarrow [R(\hat{\bfy}_t)]_{1/n}$
  \Else
    \State $R_t \leftarrow R_{t-1}$
    \State $\bfl_t \leftarrow \bfl_{t-1}$
  \EndIf
  \State Release $R_t$
\EndFor
\end{algorithmic}
\end{algorithm}

As pointed out in Section 5, the Ladder is vulnerable to the step-forward Freedman's attack because it leaks too much information about the holdout data when it releases the (rounded) score of the best submission so far. A strait forward tweak of the Ladder mechanism capable to circumvent (to some extent) this issue, is to simply add some randomness to the output of the Ladder mechanism. To this end, we propose the LadderBoot algorithm (Algorithm \ref{alg:ladderbootttest}), which adopts a bootstrap based privacy mechanism in order to generate the necessary randomness. Contrary to the Ladder, which releases the rounded score whenever the condition $R(\hat{\bfy}_t) < R_{t-1} - c_\alpha \; \mbox{sd}(\bfl_t - \bfl_{t-1})/\sqrt{n}$ is met, the LadderBoot releases a bootstrapped estimate of $R(\hat{\bfy}_t)$, computed from $b$ bootstrap replications of the loss vector $\bfl_{t}$ (as detailed in lines 12 and 13) if the condition is met, and a bootstrapped estimate of $R(\hat{\bfy}_{t-1})$, otherwise (lines 15 and 16).

\begin{algorithm}
\caption{LadderBoot mechanism (based on the paired t-test statistic)}\label{alg:ladderbootttest}
\begin{algorithmic}[1]
\State \textbf{Input}: Data, $\bfD = \{ (\bfx_1, y_1), \ldots, (\bfx_n, y_n) \}$ \\
Significance level $\alpha$ for underlying paired t-test, or corresponding critical value $c_\alpha = F^{-1}_{n-1}(\alpha)$ \\
Number of bootstrap replications $b$ \\
Assign initial estimate $R_0 \leftarrow \infty$ \\
Assign initial loss vector $l_0 = (0)_{i=1}^{n}$.
\For{$t = 1, 2, \ldots, t_{max}$}
  \State Receive prediction $\hat{\bfy}_t$
  \State Compute loss vector $\bfl_t \leftarrow (l(\hat{y}_{i} \, , \, y_i))_{i=1}^n$
  \State Compute $R(\hat{\bfy}_t) \leftarrow n^{-1} \sum_{i=1}^{n} l_{t,i}$
  \State Compute $R(\hat{\bfy}_{t-1}) \leftarrow n^{-1} \sum_{i=1}^{n} l_{t-1,i}$
  \If{$R(\hat{\bfy}_t) < R(\hat{\bfy}_{t-1}) - c_\alpha \; \mbox{sd}(\bfl_t - \bfl_{t-1})/\sqrt{n}$}
    \State Generate $b$ bootstrap replications $\bfl_t^\ast$ of $\bfl_t$, and for each compute $R^{\ast}_{j,t} = n^{-1} \sum_{i=1}^{n} l_{t,i}^\ast$
    \State $R_t^\ast \leftarrow b^{-1} \sum_{j = 1}^{b} R_{j,t}^{\ast}$
  \Else
    \State Generate $b$ bootstrap replications $\bfl_{t-1}^\ast$ of $\bfl_{t-1}$, and for each compute $R^{\ast}_{j,t-1} = n^{-1} \sum_{i=1}^{n} l_{t-1,i}^\ast$
    \State $R_t^\ast \leftarrow b^{-1} \sum_{j = 1}^{b} R_{j,t-1}^{\ast}$
    \State $\bfl_t \leftarrow \bfl_{t-1}$
  \EndIf
  \State Release $R_t^\ast$
\EndFor
\end{algorithmic}
\end{algorithm}

In the Ladder and LadderBoot algorithms, the decision to release a new score is implicitly guided by a one-sided paired t-test of loss vectors (equation \ref{eq:pairedttest}). We point out, nonetheless, that the adoption of this implicit test restricts the applicability of the Ladder and LadderBoot mechanisms to performance metrics which correspond to empirical risk estimators, as defined in equation \ref{eq:empiricalrisk}. For instance, the mean squared error metric corresponds to the average of element-wise quadratic losses, $l_i = (\hat{y}_i - y_i)^2$, whereas the mean absolute error is given by the average of absolute losses, $l_i = |\hat{y}_i - y_i|$. In the context of classification problems, the classification error is given by the average of zero-one losses, $l_i = \ind\{\hat{y}_i \not= y_i\}$. The paired t-test cannot, however, be directly applied to commonly adopted metrics such as concordance correlation coefficient and Pearson and Spearman's correlation coefficients, which cannot be directly expressed as an average of element-wise loss values.

In order to extend the applicability of the Ladder and LadderBoot mechanisms to other performance metrics, we have to replace the paired t-test by a more general statistical procedure. Here, we adopt a Bayesian perspective, and employ the Bayesian bootstrap\cite{rubin1981} to estimate the posterior distribution of the the statistic,
$$
\Delta s = s(\hat{\bfy}_t \, , \, \bfy) - s(\hat{\bfy}_{t-1} \, , \, \bfy)~,
$$
where $s(\hat{\bfy} \, , \, \bfy)$ represents an arbitrary scoring metric, and perform a one-sided Bayesian hypothesis test to determine whether the current score is statistically better than the best score so far. For scoring metrics such as correlation coefficients, where larger scores indicate better performance, we perform the one-sided test,
$$
H_0: \Delta s \leq 0 \;\;\; \mbox{vs} \;\;\; H_1: \Delta s > 0~,
$$
whereas for scoring metrics for which smaller values indicate better performance we test,
$$
H_0: \Delta s \geq 0 \;\;\; \mbox{vs} \;\;\; H_1: \Delta s < 0~.
$$
The Bayesian test is based on the posterior odds in favor of $H_1$,
$$
PO = \frac{P(H_1 \mid D)}{P(H_0 \mid D)} = \frac{P(H_1 \mid D)}{1 - P(H_1 \mid D)},
$$
where the posterior probability of the alternative hypothesis, $P(H_1 \mid D)$, is given by the proportion of bootstrap samples (in the Bayesian bootstrap posterior distribution) for which $H_1$ is true (i.e., for which $\Delta s > 0$, in the ``the larger, the better" case, and for which $\Delta s < 0$, in the ``the smaller, the better" case). We call these Bayesian bootstrap variations of the Ladder and LadderBoot algorithms the BayesBootLadder (Algorithm \ref{alg:bayesbootladder}) and BayesBootLadderBoot (Algorithm \ref{alg:bayesbootladderboot}) mechanisms, respectively.

\begin{algorithm}
\caption{BayesBootLadder mechanism (for ``the larger, the better" scoring metrics)}\label{alg:bayesbootladder}
\begin{algorithmic}[1]
\State \textbf{Input}: Data, $\bfD = \{ (\bfx_1, y_1), \ldots, (\bfx_n, y_n) \}$ \\
Number of bootstrap replications, $B$, for $\Delta s$ distribution \\
Number of bootstrap replications, $b$, for released score \\
Posterior odds threshold $PO_{thr}$ \\
Assign initial prediction, $\hat{\bfy}_0$ \\
Assign initial score, $s(\hat{\bfy}_0 \, , \, \bfy)$.
\For{$t = 1, 2, \ldots, t_{max}$}
  \State Receive prediction $\hat{\bfy}_t$
  %\State Compute score $s_t(\hat{\bfy}_t \, , \, \bfy)$
  \State Generate $B$ paired bootstrap replications of the statistic $\Delta s = s(\hat{\bfy}_t \, , \, \bfy) - s(\hat{\bfy}_{t-1} \, , \, \bfy)$
  \State Compute the posterior odds, $PO$, for the Bayesian hypothesis test, $H_0: \Delta s \leq 0$ vs $H_1: \Delta s > 0$.
  \If{$PO \geq PO_{thr}$}
    \State $s_t \leftarrow [s(\hat{\bfy}_t \, , \, \bfy)]_{1/n}$
  \Else
    \State $s_{t} \leftarrow [s(\hat{\bfy}_{t-1} \, , \, \bfy)]_{1/n}$
    \State $\hat{\bfy}_t \leftarrow \hat{\bfy}_{t-1}$
  \EndIf
  \State Release $s_t$
\EndFor
\end{algorithmic}
\end{algorithm}

\begin{algorithm}
\caption{BayesBootLadderBoot mechanism (for ``the larger, the better" scoring metrics)}\label{alg:bayesbootladderboot}
\begin{algorithmic}[1]
\State \textbf{Input}: Data, $\bfD = \{ (\bfx_1, y_1), \ldots, (\bfx_n, y_n) \}$ \\
Number of bootstrap replications, $B$, for $\Delta s$ distribution \\
Number of bootstrap replications, $b$, for released score \\
Posterior odds threshold $PO_{thr}$ \\
Assign initial prediction, $\hat{\bfy}_0$ \\
Assign initial score, $s(\hat{\bfy}_0 \, , \, \bfy)$.
\For{$t = 1, 2, \ldots, t_{max}$}
  \State Receive prediction $\hat{\bfy}_t$
  \State Generate $B$ paired bootstrap replications of the statistic $\Delta s = s(\hat{\bfy}_t \, , \, \bfy) - s(\hat{\bfy}_{t-1} \, , \, \bfy)$
  \State Compute the posterior odds, $PO$, for the Bayesian hypothesis test, $H_0: \Delta s \leq 0$ vs $H_1: \Delta s > 0$.
  \If{$PO \geq PO_{thr}$}
    \State Generate $b$ bootstrap replications $s_{j,t}^{\ast}$ of the statistic $s_t$
    \State $s_t^\ast \leftarrow b^{-1} \sum_{j = 1}^{b} s_{j,t}^{\ast}$
  \Else
    \State Generate $b$ bootstrap replications $s_{j,t-1}^{\ast}$ of the statistic $s_{t-1}$
    \State $s_{t}^\ast \leftarrow b^{-1} \sum_{j = 1}^{b} s_{j,t-1}^{\ast}$
    \State $\hat{\bfy}_t \leftarrow \hat{\bfy}_{t-1}$
  \EndIf
  \State Release $s_t^\ast$
\EndFor
\end{algorithmic}
\end{algorithm}

Computation of the posterior distribution of the $\Delta s$ statistic (line 9 of Algorithms \ref{alg:bayesbootladder} and \ref{alg:bayesbootladderboot}) with the Bayesian bootstrap is based on weighting the $(\hat{\bfy}_t, \hat{\bfy}_{t-1}, \bfy)$ data according to weights sampled from a Dirichlet distribution. For instance, suppose we are interested in using Pearson's correlation (COR) as a performance metric. In order to sample one data point from the posterior distribution of $\Delta\mbox{COR} = \mbox{COR}(\hat{\bfy}_t \, , \, \bfy) - \mbox{COR}(\hat{\bfy}_{t-1} \, , \, \bfy)$, we only need to sample a weight vector, $\bfw = (w_1, \ldots, w_n)^t$ from a $\bfw \sim \mbox{Dirichlet}_{n}(1, \ldots, 1)$ distribution and compute,
\begin{align}
\Delta\mbox{COR} &= \dfrac{\sum_i w_i \, y_i \, \hat{y}_{t,i} \, - \, (\sum_{i} w_i \, y_i) (\sum_{i} w_i \, \hat{y}_{t,i})}{\sqrt{(\sum_i w_i \, y_i^2 - (\sum_{i} w_i \, y_i)^2) \, (\sum_i w_i \, \hat{y}_{t,i}^2 - (\sum_{i} w_i \, \hat{y}_{t,i})^2)}} \nonumber \\
&- \dfrac{\sum_i w_i \, y_i \, \hat{y}_{t-1,i} \, - \, (\sum_{i} w_i \, y_i) (\sum_{i} w_i \, \hat{y}_{t-1,i})}{\sqrt{(\sum_i w_i \, y_i^2 - (\sum_{i} w_i \, y_i)^2) \, (\sum_i w_i \, \hat{y}_{t-1,i}^2 - (\sum_{i} w_i \, \hat{y}_{t-1,i})^2)}}~. \nonumber
\end{align}
(Observe that in order to generate paired bootstrap replications, we use the same $\bfw$ vector to weight the $(\hat{\bfy}_t, \bfy)$ and $(\hat{\bfy}_{t-1}, \bfy)$ data.)

One practical constraint of the Bayesian bootstrap is that it is only readily applicable to scoring metrics which can be expressed as functions of sample moments (since the generation of the posterior distribution is based on weighting the observed data, according to Dirichlet weights as illustrated above). Although several commonly used performance metrics satisfy this constraint (e.g, mean squared error, mean absolute error, correlation coefficients, concordance correlation coefficient, classification error, and etc), there are useful metrics such as the probabilistic concordance index\cite{costello2014} which cannot be expressed as a function of sample moments.

We point out, however, that for scoring metrics not amenable to the application of the Bayesian bootstrap, we can still generate a bootstrap distribution for the $\Delta s$ statistic, using the standard non-parametric (frequentist) bootstrap using data re-sampling, and interpret it as a Bayesian posterior distribution. (It is known that the sampling distribution generated by the non-parametric bootstrap closely approximates the posterior distribution of the quantity of interest generated by the Bayesian bootstrap \cite{rubin1981}. As pointed out in Section 8.4 of \cite{eosl2001}, inferences derived from the non-parametric bootstrap can be interpreted as a ``poor man's" automatic Bayesian analysis based on a non-informative prior.)

Finally, Algorithm \ref{alg:stepforwardattack} presents in detail the step-forward Freedman's attack used to illustrate the vulnerability of the Ladder mechanism.

\begin{algorithm}
\caption{Step-forward Freedman's attack}\label{alg:stepforwardattack}
\begin{algorithmic}[1]
\State Create a set of selected features, $Selected$, and assign $Selected \, \leftarrow \, \emptyset$.
\State Create a set of remaining features, $Remain$, and assign all features to $Remain$.
\For{$i = 1, 2, \ldots, i_{max}$}
  \State For each one of the features in the remain set, $x_{rem}$, fit a separate regression model of the response on $x_{rem}$ and on the features in the $Selected$ set, and submit the models to the public leaderboard.
  \State Fit separate univariate regression models of the response on each of the features in $Remain$, and submit the models to the public leaderboard.
  \State Determine the feature, $x_{sel}$, responsible for the last jump in public leaderboard scores released by the Ladder mechanism.
  \State Update the set of selected features, $Selected \, \leftarrow \, Selected \, \cup \, x_{sel}$.
  \State Update the set of remaining features, $Remain \, \leftarrow \, Remain \, \setminus \, x_{sel}$.
\EndFor
\end{algorithmic}
\end{algorithm}

\section*{Acknowledgements}

We would like to thank Moritz Hardt for helpful comments on an earlier version of the paper.

\bibliography{template}

\clearpage

\section*{Supplementary materials}

\section*{Supplementary figures} $ $

\begin{supplefig}[!h]
\begin{center}
\includegraphics[angle=270, scale = 0.67, clip]{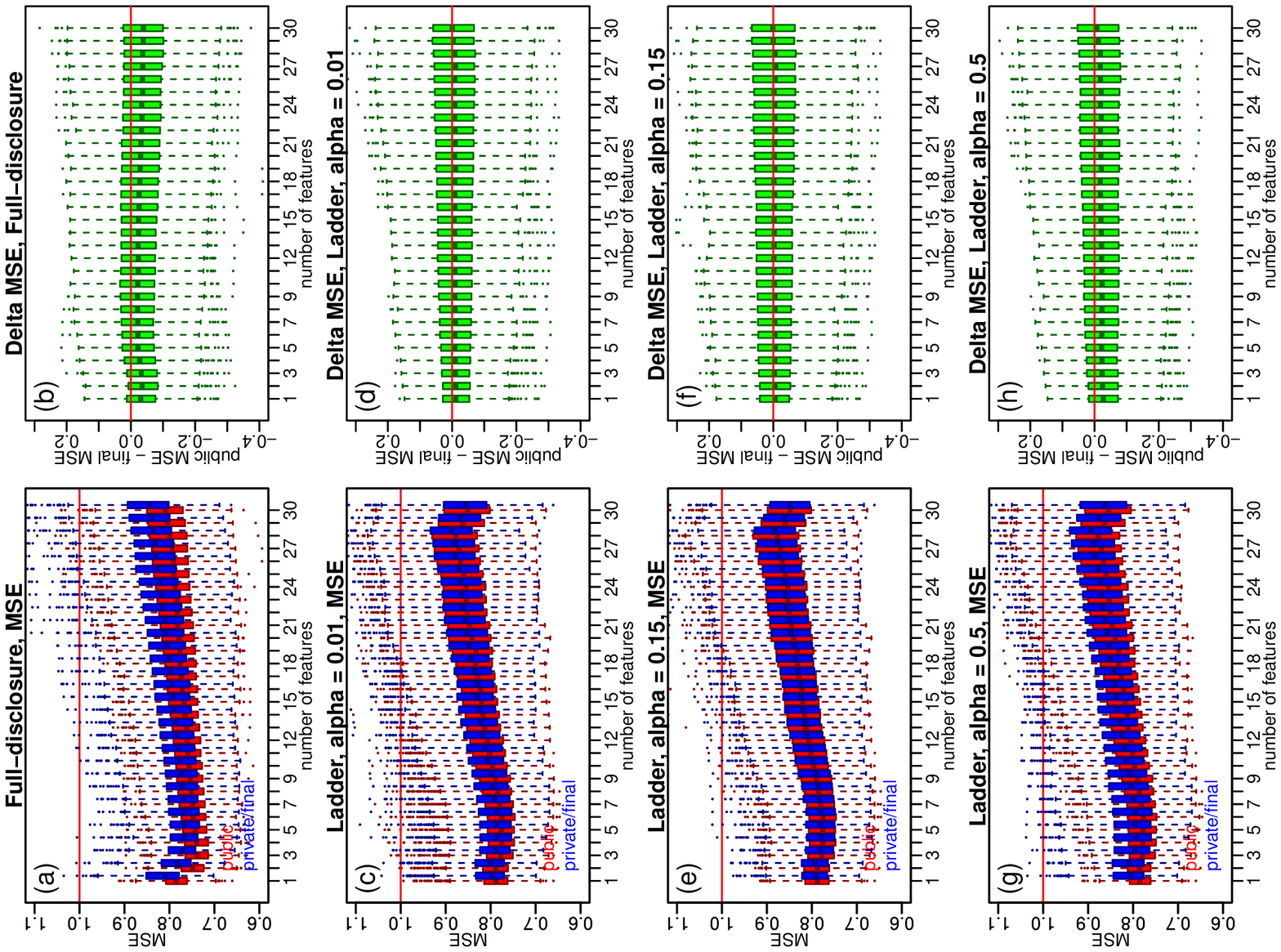}
\caption{Comparison of the Full-disclosure vs Ladder mechanisms (based on $\alpha$ equal to 0.01, 0.15, and 0.5) using the AD Challenge data. Results based on 1,000 random data splits, and on a subset of 300 randomly chosen features. The Ladder mechanism seems to provide a reduction in over-fitting but, interestingly, it tended to work slightly better with $\alpha = 0.15$ (panels e and f), whereas in the permuted responses experiment $\alpha = 0.01$ was best (Fig.\ref{fig:standarvsladderpermuted}c and d in the main text).}
\label{fig:standardvsladderoriginal}
\end{center}
\end{supplefig}

%\clearpage

\begin{supplefig}[!h]
\begin{center}
\includegraphics[angle=270, scale = 0.47, clip]{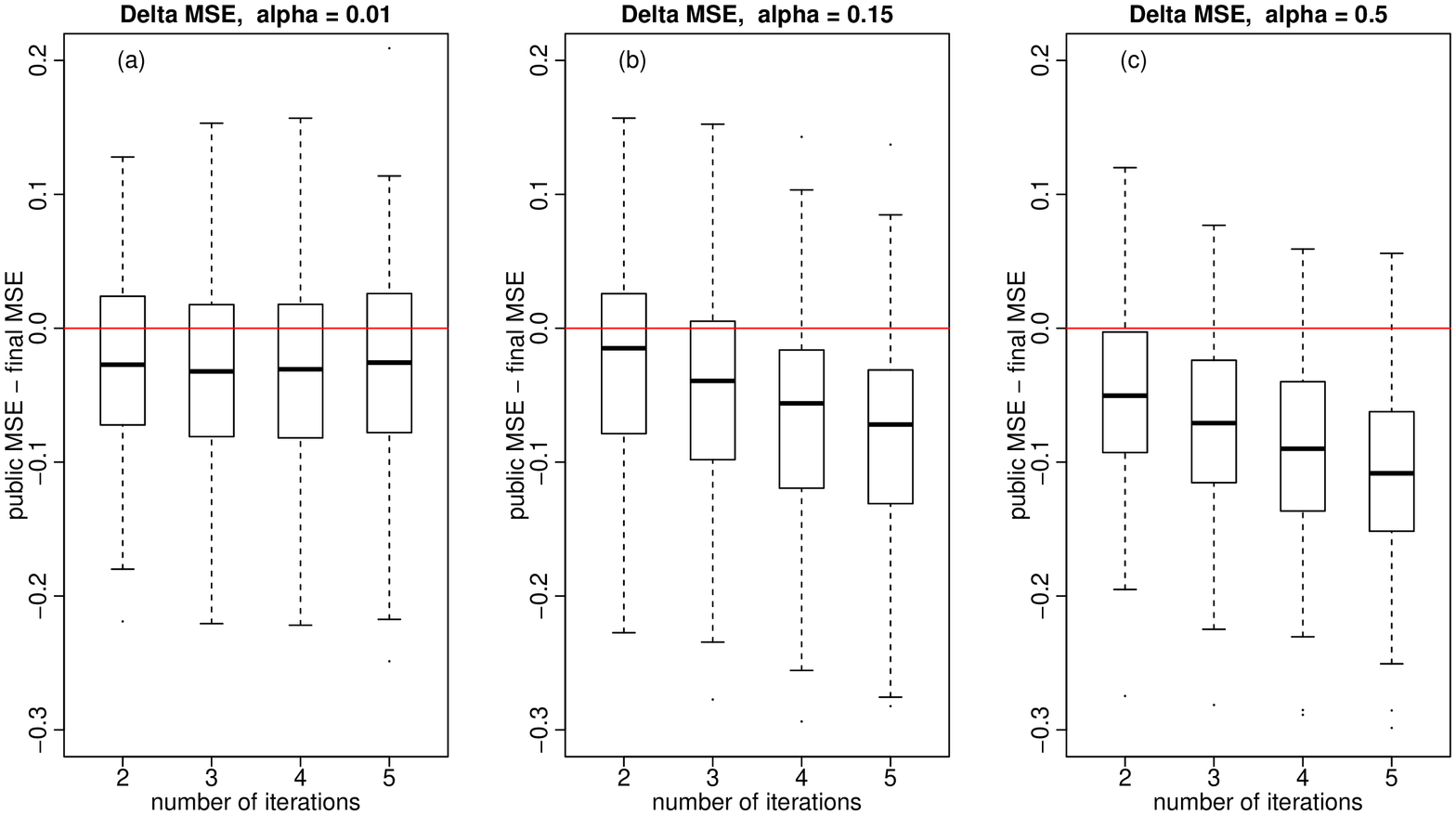}
\caption{Over-fitting incurred by the Ladder (based on $\alpha$ equal to 0.01, 0.15, and 0.5) mechanism under step-forward attacks with increasing number of iterations, using the AD Challenge data. Results based on 100 random data splits, and on a subset of 300 randomly chosen features.}
\label{fig:stepforwardladderoriginal}
\end{center}
\end{supplefig}

%\clearpage

\begin{supplefig}[!h]
\begin{center}
\includegraphics[angle=270, scale = 0.6, clip]{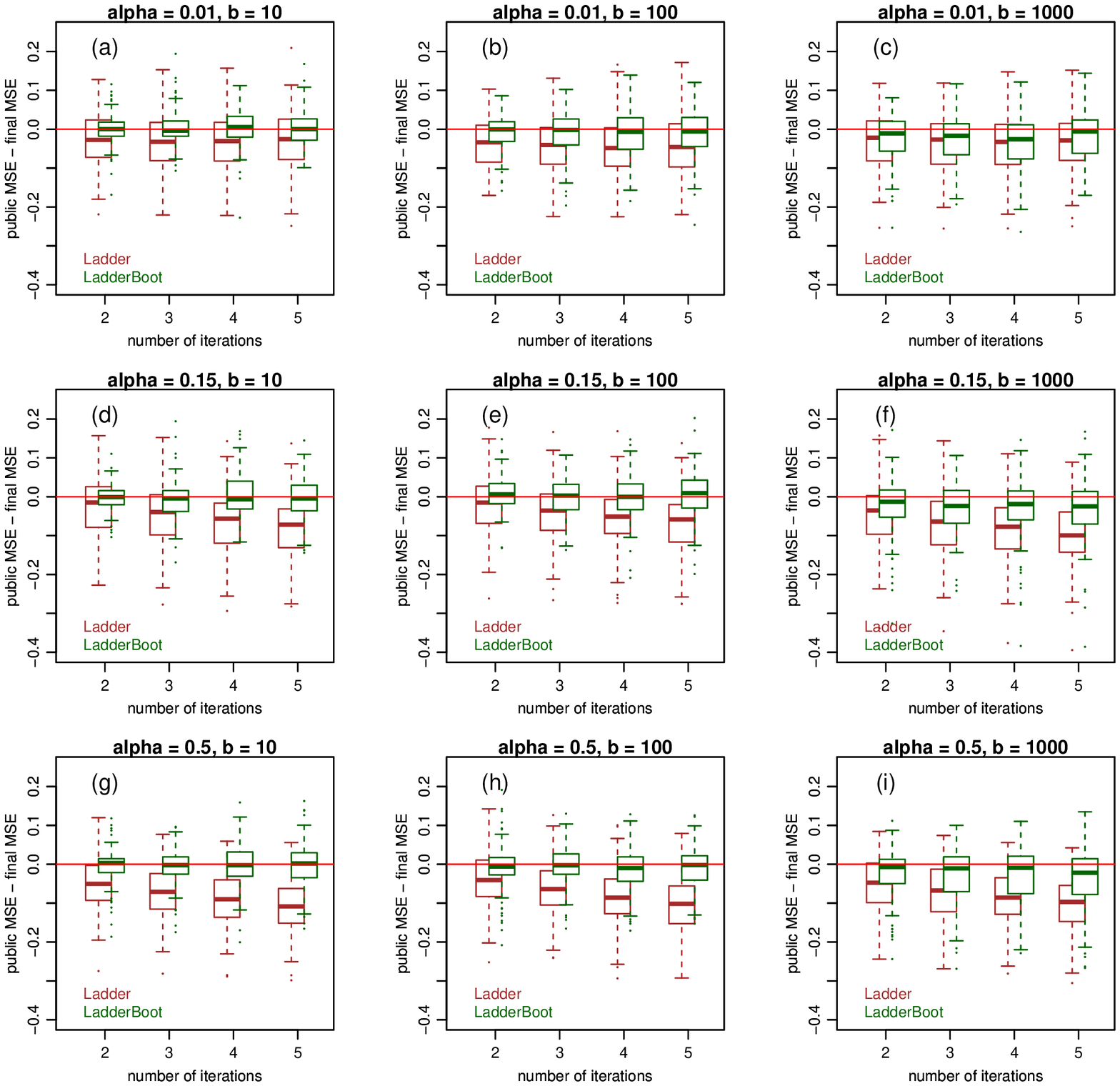}
\caption{Comparison of the $\Delta$MSE values released by the Ladder and LadderBoot algorithms, for all combinations of $\alpha = \{ 0.01, 0.15, 0.5 \}$ and $b = \{ 10, 100, 1000\}$, using the AD Challenge data. Results based on 100 random data splits, and on 300 randomly selected features.}
\label{fig:fig9}
\end{center}
\end{supplefig}

\clearpage

\begin{supplefig}[!h]
\begin{center}
\includegraphics[angle=270, scale = 0.6, clip]{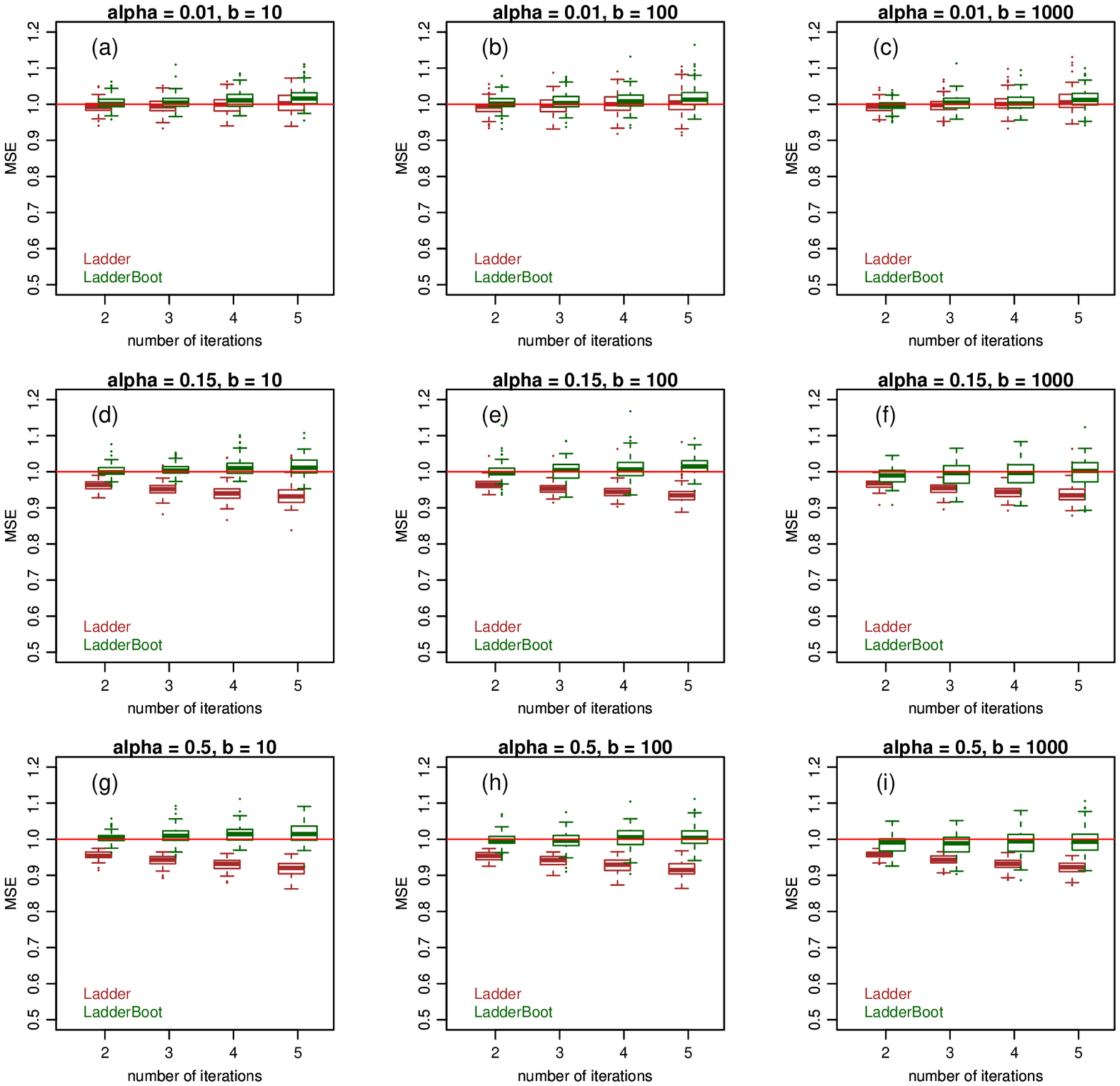}
\caption{Comparison of the MSE scores released by the Ladder and LadderBoot algorithms, for all combinations of $\alpha = \{ 0.01, 0.15, 0.5 \}$ and $b = \{ 10, 100, 1000\}$, using the AD Challenge data. Results based on 100 permutations of the response data, and on 300 randomly selected features. Note that MSE values close to 1 indicate a small amount of over-fitting.}
\label{fig:fig10}
\end{center}
\end{supplefig}

\clearpage

\begin{supplefig}[!h]
\begin{center}
\includegraphics[angle=270, scale = 0.6, clip]{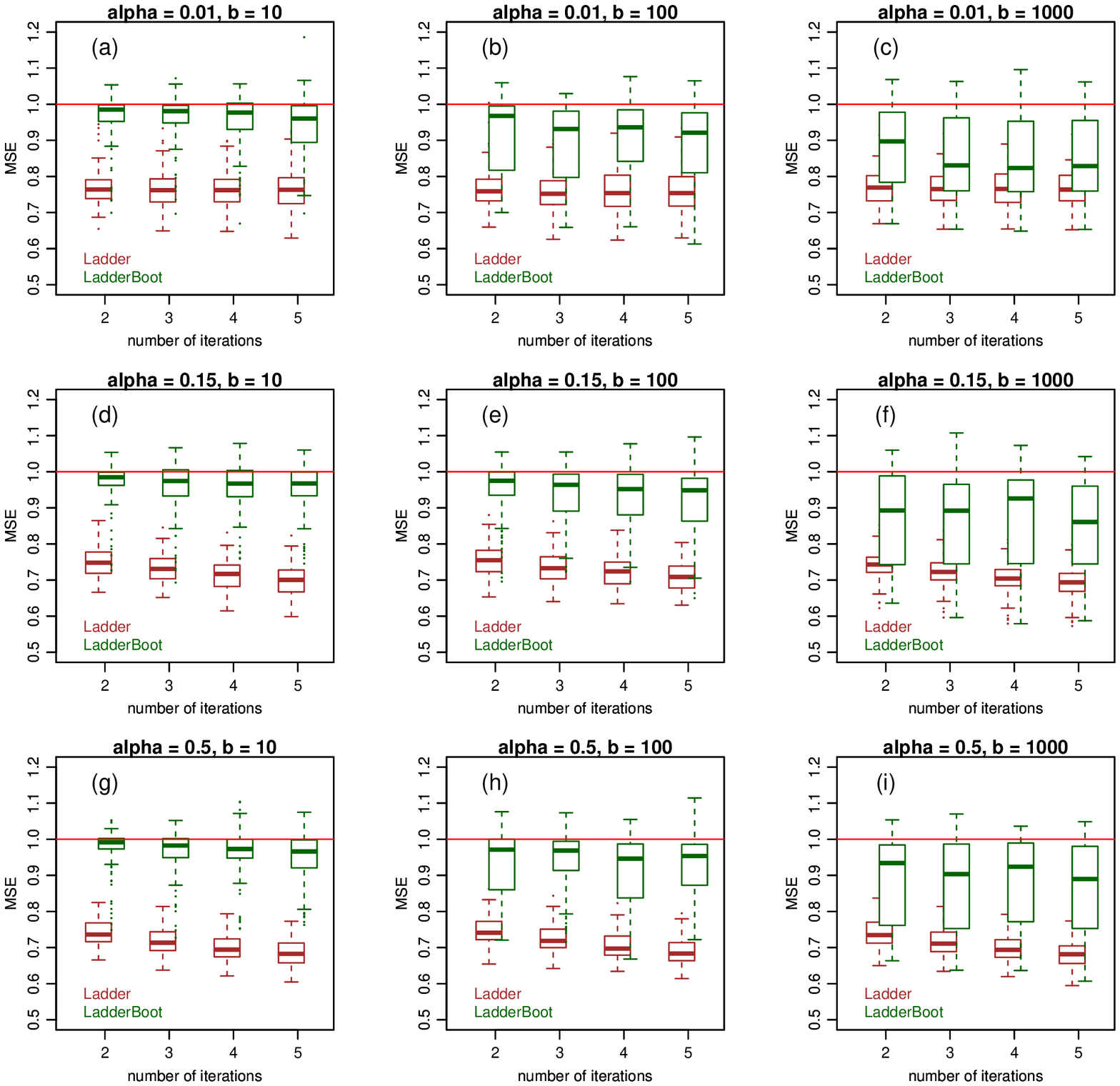}
\caption{Comparison of the MSE scores released by the Ladder and LadderBoot algorithms, for all combinations of $\alpha = \{ 0.01, 0.15, 0.5 \}$ and $b = \{ 10, 100, 1000\}$, using the AD Challenge data. Results based on 100 random data splits, and on 300 randomly selected features. In all cases, we observe larger MSE scores for the LadderBoot mechanism. Also, the LadderBoot scores tended to show larger variability than the Ladder for larger values of $b$ (note the larger spread of the dark green boxplots as $b$ increases from 10 to 1000). While the larger MSE scores suggest that, in general, the bootstrapped values released by the LadderBoot make it harder to rank and select the features showing the strongest associations to the response, the larger spreads suggest that the amount of ``bootstrap noise" was not enough in some replications of the experiments which adopted larger $b$ values.}
\label{fig:fig11}
\end{center}
\end{supplefig}

\clearpage
\section*{Supplementary Text S1: Comparison of the Ladder vs LadderBoot mechanisms under Freedman's attack}

In the main text we illustrated how the LadderBoot mechanism leads to more robust public leaderboards under the step-forward attack in comparison to the Ladder mechanism. Here, we evaluate its performance under the less aggressive Freedman's attack. Figures S\ref{fig:laddervsladderbootMSEpermuted} and S\ref{fig:laddervsladderbootMSEoriginal} report, respectively, the public and private leaderboard scores for permuted responses and random data split experiments using the AD challenge data, for all combinations of the $\alpha = \{0.01, 0.15, 0.5 \}$ and $b = \{ 10, 100, 1000 \}$ parameters.

\begin{supplefig}[!h]
\begin{center}
\includegraphics[angle=270, scale = 0.7, clip]{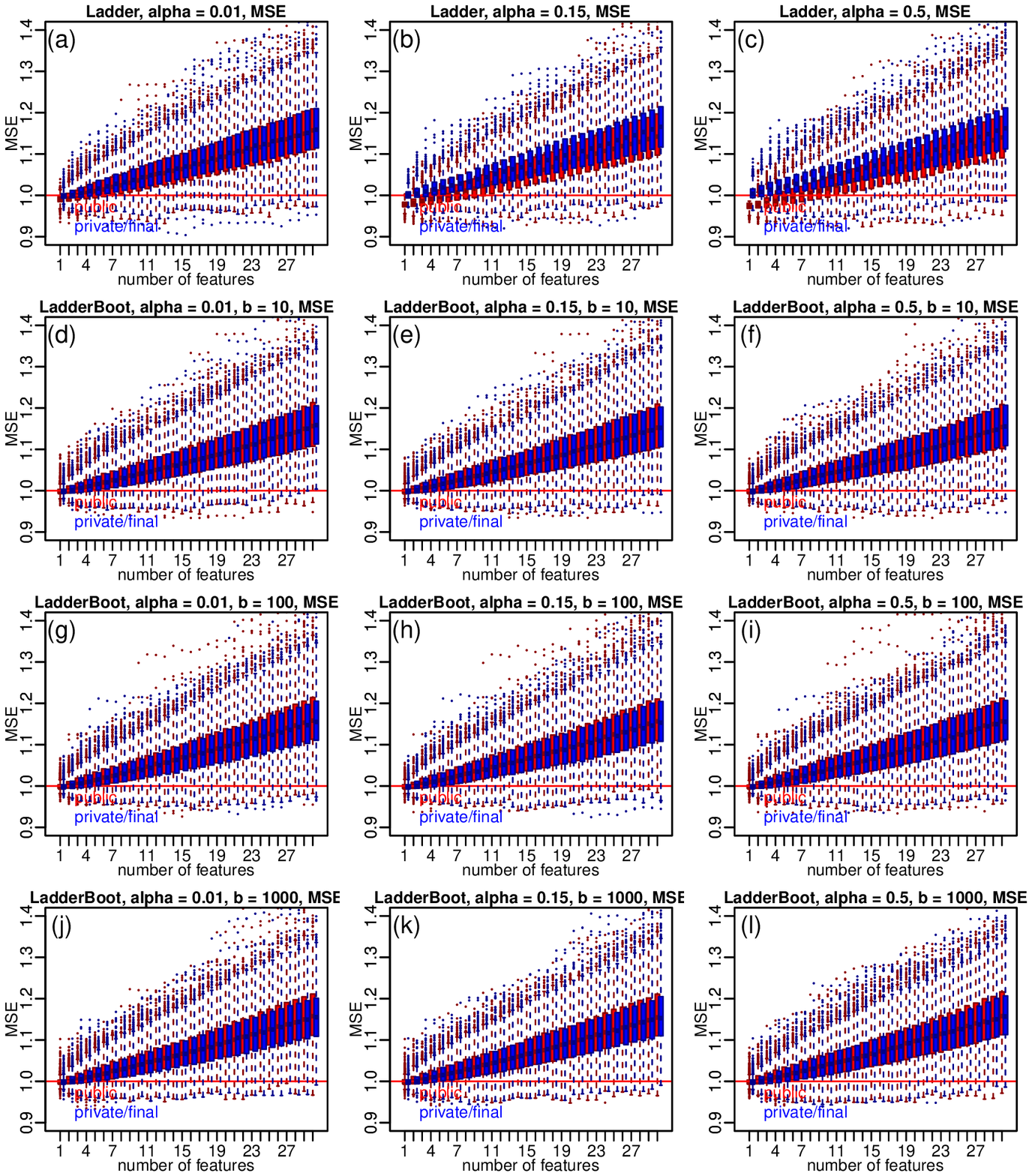}
\caption{Comparison of the Ladder and LadderBoot mechanisms under Freedman's attack, using the AD Challenge data. Results based on 1,000 permutations of the response data, and on 300 randomly selected features.}
\label{fig:laddervsladderbootMSEpermuted}
\end{center}
\end{supplefig}

\begin{supplefig}[!h]
\begin{center}
\includegraphics[angle=270, scale = 0.7, clip]{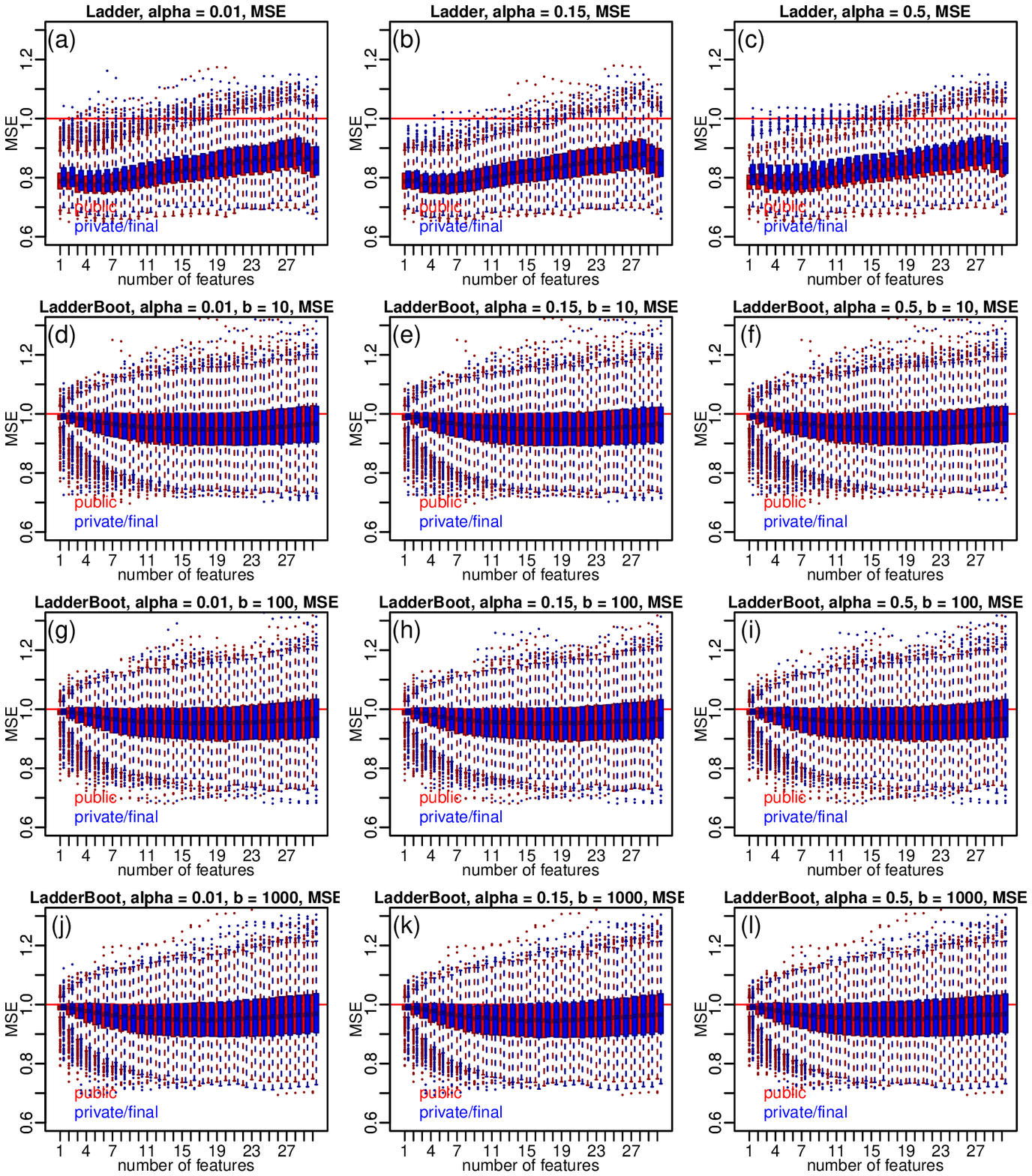}
\caption{Comparison of the Ladder and LadderBoot mechanisms under Freedman's attack, using the AD Challenge data. Results based on 1,000 random data splits, and on 300 randomly selected features.}
\label{fig:laddervsladderbootMSEoriginal}
\end{center}
\end{supplefig}

Overall, Figure S\ref{fig:laddervsladderbootMSEpermuted} reports comparable performances of both mechanisms on most cases, but with the LadderBoot mechanism slightly outperforming the Ladder algorithm for $\alpha = 0.15$ (compare panel b against panels e, h, and k) and $\alpha = 0.5$ (compare panel c against panels f, i, and l).

Figure S\ref{fig:laddervsladderbootMSEoriginal} shows some interesting patterns. First, both Ladder and LadderBoot mechanisms seemed to perform well in general (i.e., the the red and blue boxplots tend to be close in all settings). Second, while the Ladder mechanism tended to work best with $\alpha = 0.15$ (compare panel b against panels a and c), the LadderBoot performed uniformly well across all $\alpha$ levels (as measured by the closeness of the red and blue boxplots). Third, the MSE scores of the models built according to the LadderBoot released scores tended to be larger (and more variable, for larger $k$) than the MSE scores of the models built from the Ladder output.

This third observation is likely a consequence of the fact that the bootstrapped values released by the LadderBoot make it harder to rank the univariate features showing the strongest associations to the response. Of course, this does not mean that the LadderBoot prevents a participant from finding a good model (i.e., with low MSE score), it only shows that it is hard to do so by guiding the model refinement activities on the scores released by the public leaderboard (illustrating the increased effectiveness of the LadderBoot mechanism against the Freedman's attack). Clearly, if a participant adopts best practices for avoiding over-fitting to the training (and to the public leaderboard) and generates a sensible predictive model with good generalization performance, the MSE score released by the LadderBoot algorithm will be low. Furthermore, as we illustrate in Supplementary Text S4, the score will likely be close to the public leaderboard score since the higher the predictive performance of a submission, the smaller is the variability of the LadderBoot released scores.

\clearpage
\section*{Supplementary Text S2: BayesBootLadder and BayesBootLadderBoot algorithms: an illustrative example}

We illustrate the application of the BayesBootLadder and BayesBootLadderBoot algorithms to the AD challenge data, using Lin's concordance correlation coefficient. We compare the performances of the Full-disclosure, BayesBootLadder, and BayesBootLadderBoot under Freedman's attack. For consistency with the $\alpha$ values employed in the previous experiments we adopt posterior odds thresholds of $\{99, 5.67, 1\}$, corresponding to $(1 - \alpha)/\alpha$ for $\alpha = \{0.01, 0.15, 0.5\}$ (the connection between p-values and posterior odds follows from the fact that the posterior probability of the null hypothesis in the one sided hypothesis test based on the Bayesian bootstrap closely approximates the p-value from a one-sided test based on the frequentist bootstrap\cite{dreamscoring2014}). As before, we perform both permuted responses and random data split experiments. However, because the Bayesian bootstrap approaches require an increased amount of computation we performed only 100 replications of each experiment.

Figure S\ref{fig:cccstandard} presents the results from the permuted responses (panels a and c) and random data splits (panels b and d) experiments for the Full-disclosure mechanism. Similarly to our previous results, we observe strong over-fitting in the permuted responses case, while only a moderate amount for the random data splits experiment.

\begin{supplefig}[!h]
\begin{center}
\includegraphics[angle=270, scale = 0.64, clip]{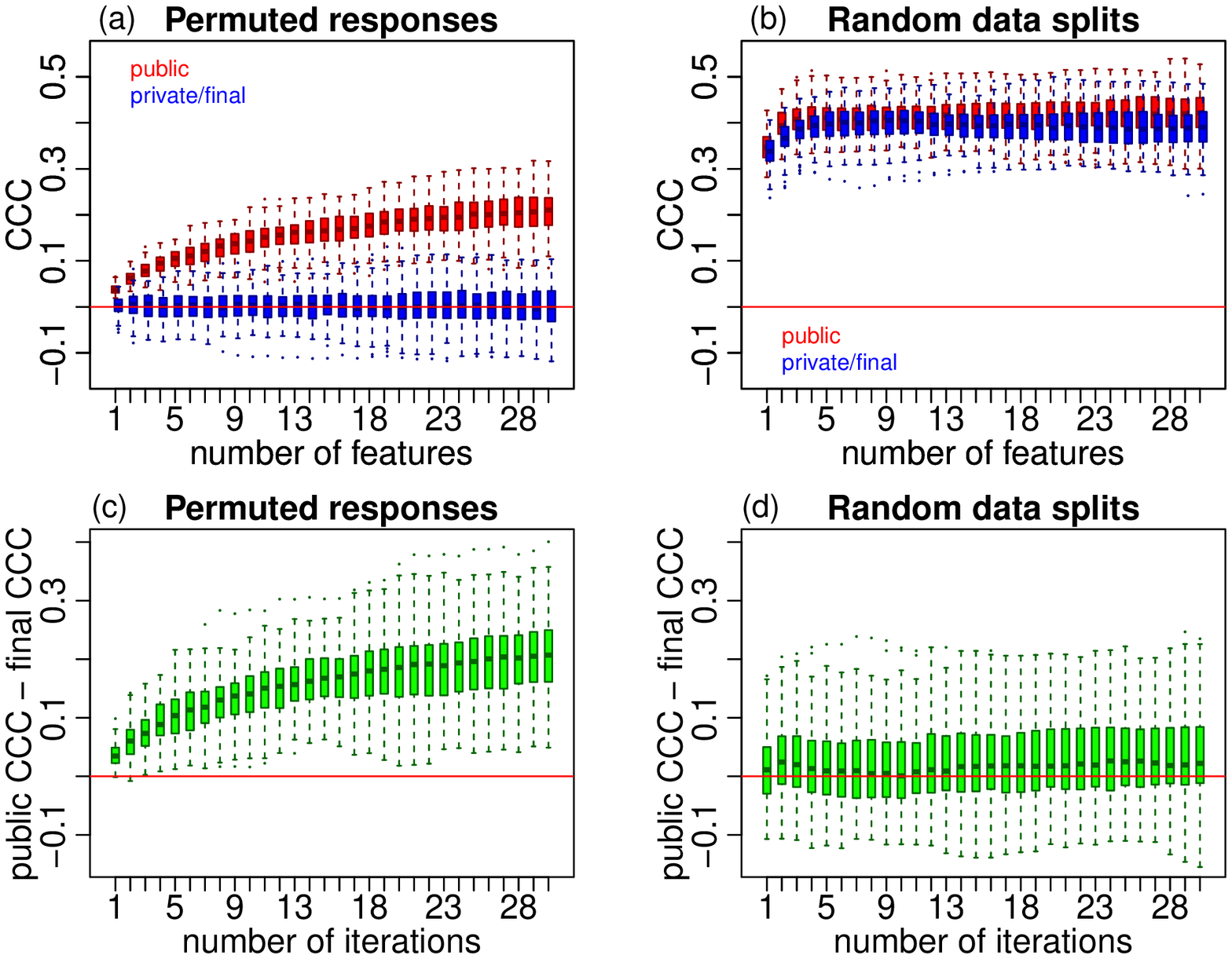}
\caption{Full-disclosure mechanism results, based on Lin's concordance correlation coefficient, for the permuted responses and random data split experiments under Freedman's attack. Results based on 100 replications, and on 300 randomly selected features.}
\label{fig:cccstandard}
\end{center}
\end{supplefig}

\begin{supplefig}[!h]
\begin{center}
\includegraphics[angle=270, scale = 0.7, clip]{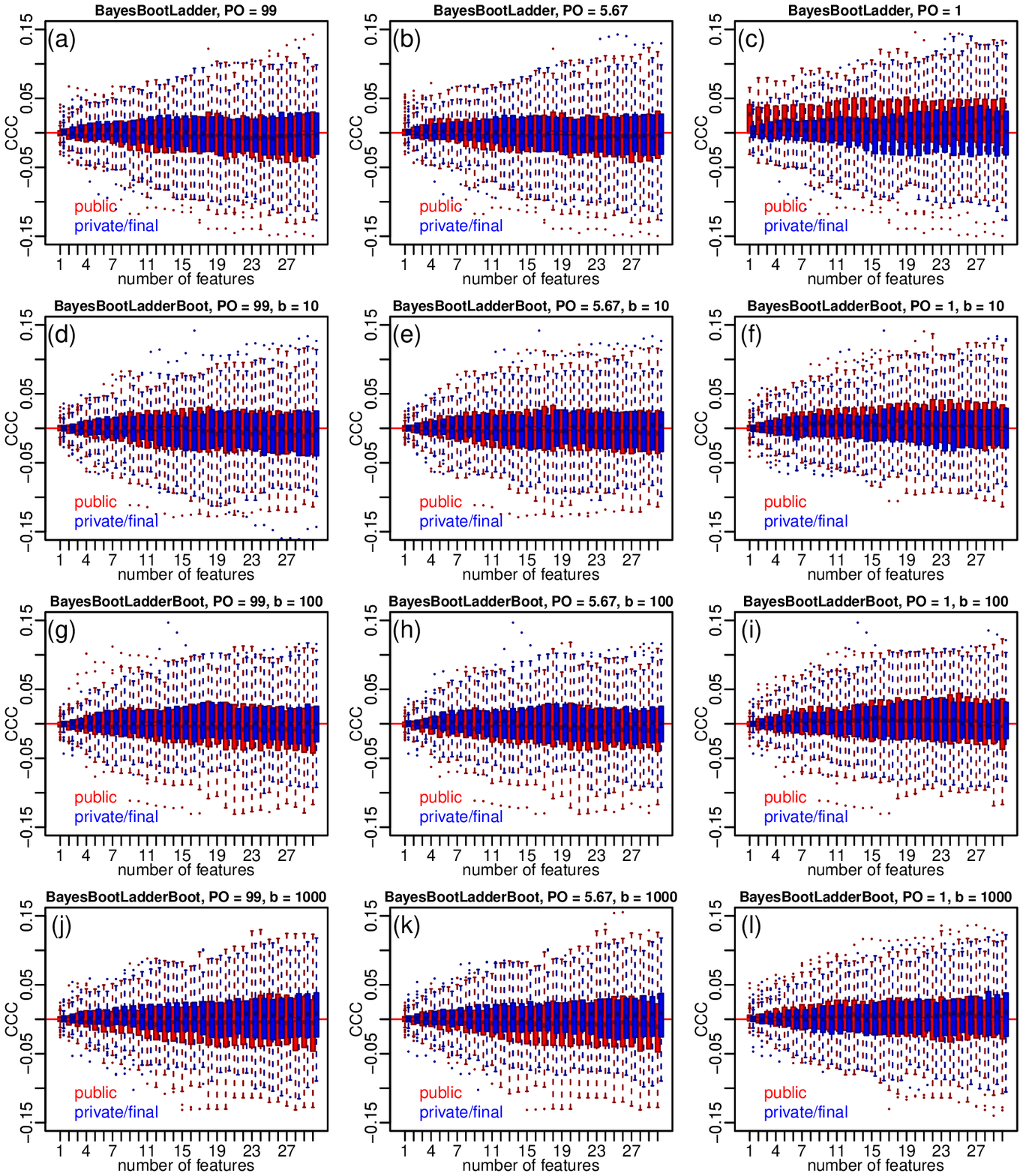}
\caption{Comparison of the BayesBootLadder and BayesBootLadderBoot mechanisms, based on Lin's concordance correlation coefficient, under Freedman's attack. Results based on 100 permutations of the response data, and on 300 randomly selected features.}
\label{fig:laddervsladderbootCCCpermuted}
\end{center}
\end{supplefig}

Figures S\ref{fig:laddervsladderbootCCCpermuted} and S\ref{fig:laddervsladderbootCCCoriginal} report the results for the BayesBootLadder and BayesBootLadderBoot mechanisms for the permuted responses and random data split experiments, respectively. The public leaderboard scores released by the BayesBootLadder and BayesBootLadderBoot algorithms seem to track well with the private leaderboard scores, except, maybe, for panels c, j, and k in Figure S\ref{fig:laddervsladderbootCCCpermuted}. We point out, however, that this observation and the generally noisier appearance of the boxplots on Figures S\ref{fig:laddervsladderbootCCCpermuted} and S\ref{fig:laddervsladderbootCCCoriginal}, when compared to Figures S\ref{fig:laddervsladderbootMSEpermuted} and S\ref{fig:laddervsladderbootMSEoriginal}, is likely due to the fact that the results were based on only 100 replications instead of the 1,000 replications used in generation of Figures S\ref{fig:laddervsladderbootMSEpermuted} and S\ref{fig:laddervsladderbootMSEoriginal}.

\begin{supplefig}[!h]
\begin{center}
\includegraphics[angle=270, scale = 0.7, clip]{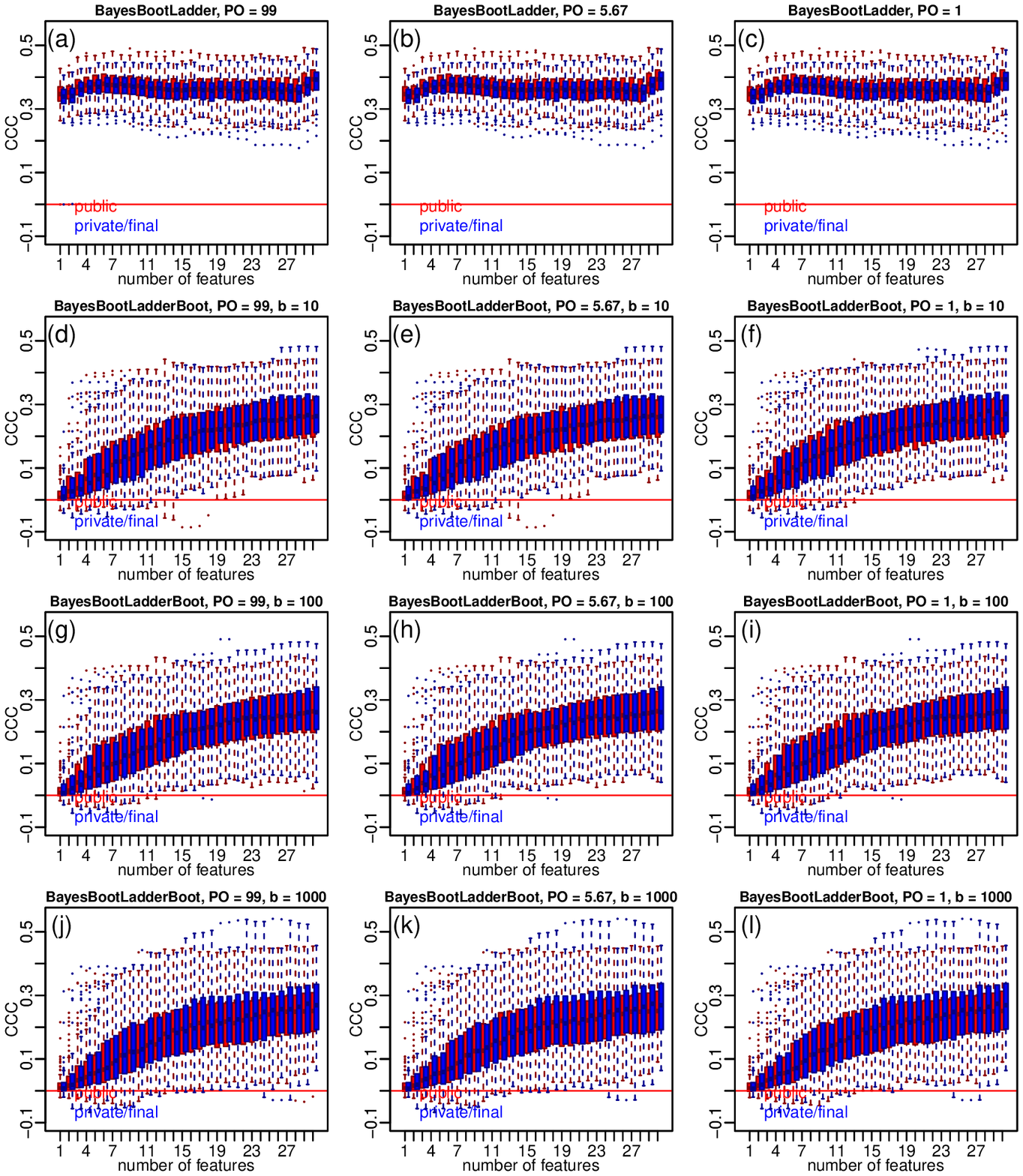}
\caption{Comparison of the BayesBootLadder and BayesBootLadderBoot mechanisms, based on Lin's concordance correlation coefficient, under Freedman's attack. Results based on 100 random data splits, and on 300 randomly selected features.}
\label{fig:laddervsladderbootCCCoriginal}
\end{center}
\end{supplefig}

\clearpage
\section*{Supplementary Text S3: Empirical determination of the number of submissions}

The key question, underlying the practical usefulness of the Ladder and LadderBoot mechanisms, is the determination of the maximum number of submissions a challenge organizer can safely allow, before participants might start over-fitting their models to the holdout data supporting the public leaderboard. Here, we present an empirical approach which can provide some guidance on the matter. We emphasize, however, that the coarse results generated by the heuristic approach we describe next can only provide a ballpark for the number of submissions and should be interpreted as a general guideline and not as a final answer (especially because the results are model dependent as illustrated in Supplementary Text S4).

Before we describe our heuristic it is important to first consider what conditions can lead to the largest amount of over-fitting generated by the step-forward attack. Clearly, the larger the number of available features, the easier it is to over-fit. However, given a fixed maximum number of submissions, the following natural question arises: is it better to run the step-forward attack for a small number of iterations over a larger subset of the features, or is it better to run it for a larger number of iterations over a smaller number of features? For concreteness, suppose we have a large number of features, $p$, but we are only allowed 120 submissions to the public leaderboard. We have a few options on how to run a step-forward attack including: ($i$) 1 iteration over 120 randomly chosen features; ($ii$) 3 iterations over 41 features, since $41 + 40 + 39 = 120$; ($iii$) 5 iterations over 26 features, since $26 + 25 + 24 + 23 + 22 = 120$; and ($iv$) 15 iterations over 15 features, since $15 + 14 + 13 + \ldots + 1 = 120$. Note that case $i$ corresponds, actually, to a simple Freedman's attack. Case $ii$ selects 3 features with non-negligible predictive ability out of a subset of 41 randomly chosen features, while case $iii$ selects 5 features out of 26 randomly chosen features. Case $iv$, on the other hand, amounts to selecting all 15 randomly chosen features.

In the context of a ``permuted responses" experiment, we would not expect to see a lot of over-fitting in case $i$, since our results suggest that the Ladder is able to withstand well the simple Freedman's attach. The same is true for case $iv$, as the selection of all 15 randomly chosen features would probably lead to a model unable to predict well the shuffled response data. Cases $ii$ and $iii$, on the other hand, have a better chance to over-fit the public leaderboard since they employ a small number of features with non-negligible ability to predict the shuffled responses. These considerations suggest that for a given fixed number of submissions, there is an optimal combination of the number of iterations and number of features that lead to a maximum amount of over-fitting by a step-forward attack. Therefore, in principle, a challenge organizer might be able to empirically determine the number of iterations needed to achieve the maximum amount of over-fitting for any fixed number of submissions.

Here, nonetheless, we employ a simplified strategy (due to computational constraints). Namely, we estimate the maximum number of submissions a challenge organizer can safely allow, by quantifying the amount of over-fitting generated by a step-forward attack examined across a grid of feature subsets of increasing size, and adopting an adaptive number of iterations corresponding to 2\% of the size of the feature subsets. The rationale for this choice goes as follows. Ideally, the number of iterations should be equal to the number of features associated to the permuted responses (since the most efficient attack, measured by the strongest amount of over-fitting, would be achieved by incorporating all features associated with the permuted responses in the multiple regression model). However, for any random sample of features, we would expect only a small number of features to be associated by chance to the permuted response data. Therefore, it seems reasonable to select the number of iterations of the step-forward attack to be proportional to a small fraction of the number of features in the random sample. Here, we arbitrary adopt the fraction to be 2\%, but recognize that the results might be affected by a different choice.

Figure S\ref{fig:numbersubmissions} reports the results of an experiment where we run the step-forward attack over 6 distinct settings, namely: 1 iteration over 50 features (50 submissions), 2 iterations over 100 features (199 submissions), 3 iterations over 150 features (447 submissions), 4 iterations over 200 features (794 submissions), 5 iterations over 250 features (1240 submissions), and 6 iterations over 300 features (1785 submissions). (Note that the number of iterations, 1, 2, 3, 4, 5, and 6, correspond to 2\% of the number of features, 50, 100, 150, 200, 250, and 300). The boxplots represent the distributions of the $\Delta$MSE score over 1,000 replications of each experiment (each replication used a distinct subset of features, randomly selected from the 2,150 available imaging features). Panels a, b, and c show the results based on $\alpha$ equal to 0.01, 0.15, and 0.5, respectively. In all cases we adopted $b = 100$ for the LadderBoot.

\begin{supplefig}[!h]
\begin{center}
\includegraphics[angle=270, scale = 0.47, clip]{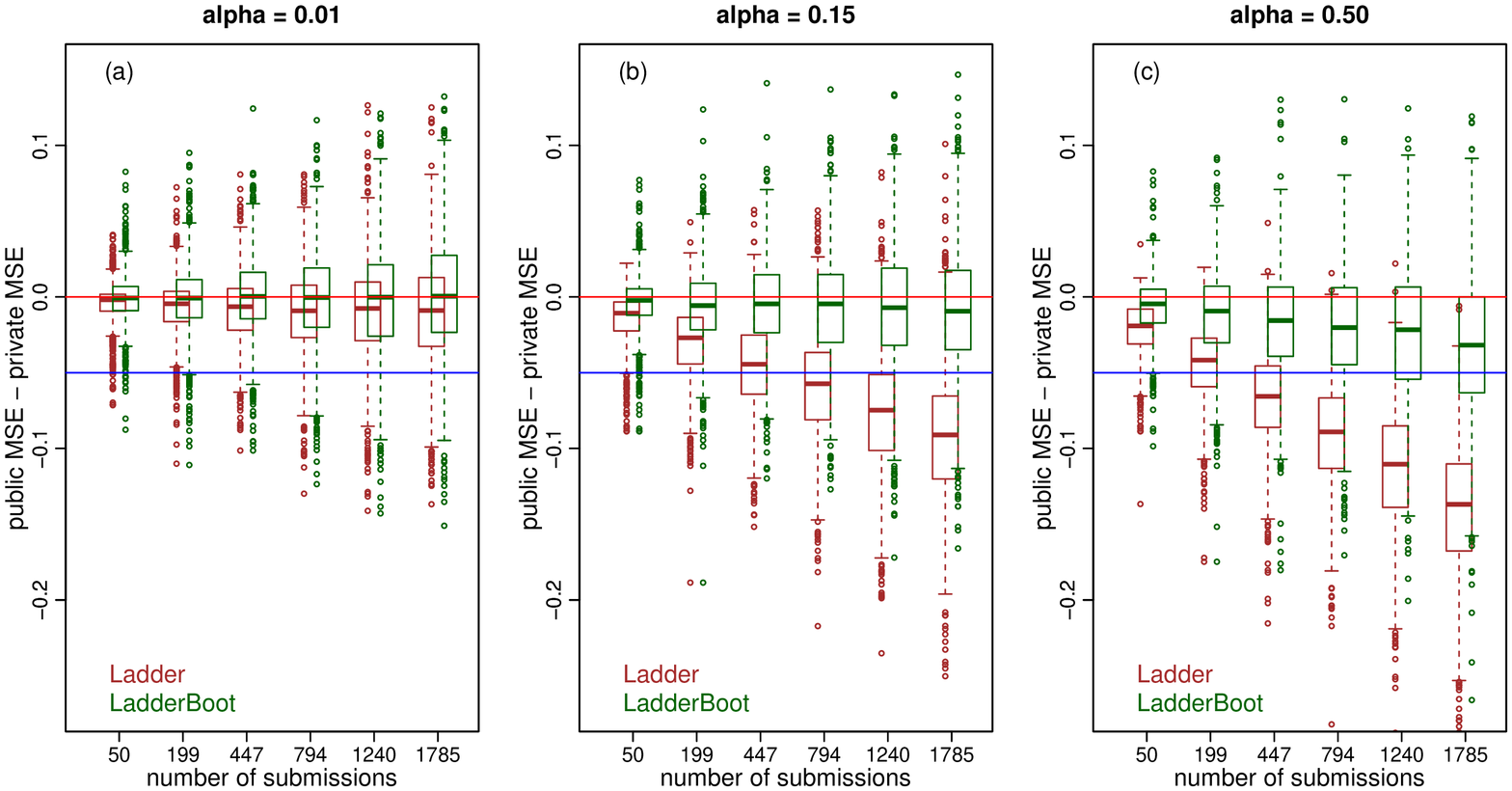}
\caption{Empirical determination of the number of submissions to the public leaderboard using the AD Challenge data. Results based on 1,000 permutations of the response data for each experimental setting.}
\label{fig:numbersubmissions}
\end{center}
\end{supplefig}

Given a fixed threshold for the amount of over-fitting that a challenge organizer is prepared to accept, it is then possible to empirically determine the maximum number of submissions that would lead to less over-fitting than the threshold. For instance, if we adopt a threshold equal to -0.05 (blue lines in Figure S\ref{fig:numbersubmissions}), we would need to restrict the number of submissions to less than 794 and 447, when adopting the Ladder mechanism with $\alpha$ equal to 0.15 and 0.5, respectively, while it would still be fine to allow 1,785 submissions when using the Ladder with $\alpha = 0.01$ or the LadderBoot irrespective of the adopted $\alpha$ value. (Note that we are employing the median of the boxplot in our comparisons to the adopted threshold.)

Figure S\ref{fig:numbersubmissions} clearly shows that the LadderBoot mechanism allows for a larger number of submissions in all settings tested, but it also suggests that adoption of $\alpha = 0.01$ can allow a large number of submissions even when adopting the Ladder leaderboard. It is important to highlight, however, that choosing a very stringent $\alpha$ level might discourage participants to engage in the challenge, as participants might be more inclined to give up if they feel their attempts to improve their models are usually unsuccessful.

We point out that it is important to perform this analysis using permuted responses, since the assumption that only a small number of features is associated with the response is not necessarily true when the connection between the response and the features is kept intact. In other words, it is harder to select a reasonable number of iterations when we run the analysis using un-shuffled response data, since we cannot easily guess the approximate number of features associated with the response due entirely to random chance.

The results presented in Figure S\ref{fig:numbersubmissions} clearly show that our rough estimates are sensible to the choice of the $\alpha$ tuning parameter employed by the Ladder and LadderBoot algorithms. So, in practice, it is left to the challenge organizers to determine a combination of $b$ and $\alpha$ (or posterior odds, when using the BayesBootLadderBoot algorithm) which would allow a reasonable number of submissions to the public leaderboard and, at the same time, would strike a good balance between protection against over-fitting and the usefulness of the released scores. In any case, we argue that even rough empirical estimates can be valuable in practice, since we might be able to scale up the number of allowed submissions, even when we are very conservative and adopt a much smaller limit than our empirical results suggest to be safe. For instance, even though the results in Figure S\ref{fig:numbersubmissions}b suggest it is probably safe to allow 1,800 submissions when we adopt the LadderBoot mechanism (based on $\alpha = 0.15$ and $b = 100$ for the AD Challenge data), the adoption of a conservative limit of 300 submissions still provides a (two orders of magnitude) increase over the usual 3 submissions, so often adopted in the DREAM challenges.

\clearpage
\section*{Supplementary Text S4: On the variability of the LadderBoot released scores}

Here, we show that the variability of the scores released by the LadderBoot mechanism is affected not only by the number of bootstraps, but also by the quality of the prediction. Higher quality predictions lead to a smaller amount of variability in the released scores when compared to lower quality predictions. To illustrate this point we show in Figure S\ref{fig:lassovsknn} the distributions of the released scores adopting $b$ equal to 10, 100, and 1000, for a prediction generated by a lasso model (panels a to c), in comparison to the released scores from a prediction generated with a k-nearest neighbors (knn) regression model adopting $k = 3$ (panels d to f). The red vertical lines correspond to the MSE score of the actual predictions (0.79 for the lasso and 1.12 for the knn). Note the consistently smaller spread of the distributions in the top panels, based on the higher quality prediction generated by the lasso, relative to the distributions generated from the knn prediction shown in the respective bottom panels.

\begin{supplefig}[!h]
\begin{center}
\includegraphics[angle=270, scale = 0.49, clip]{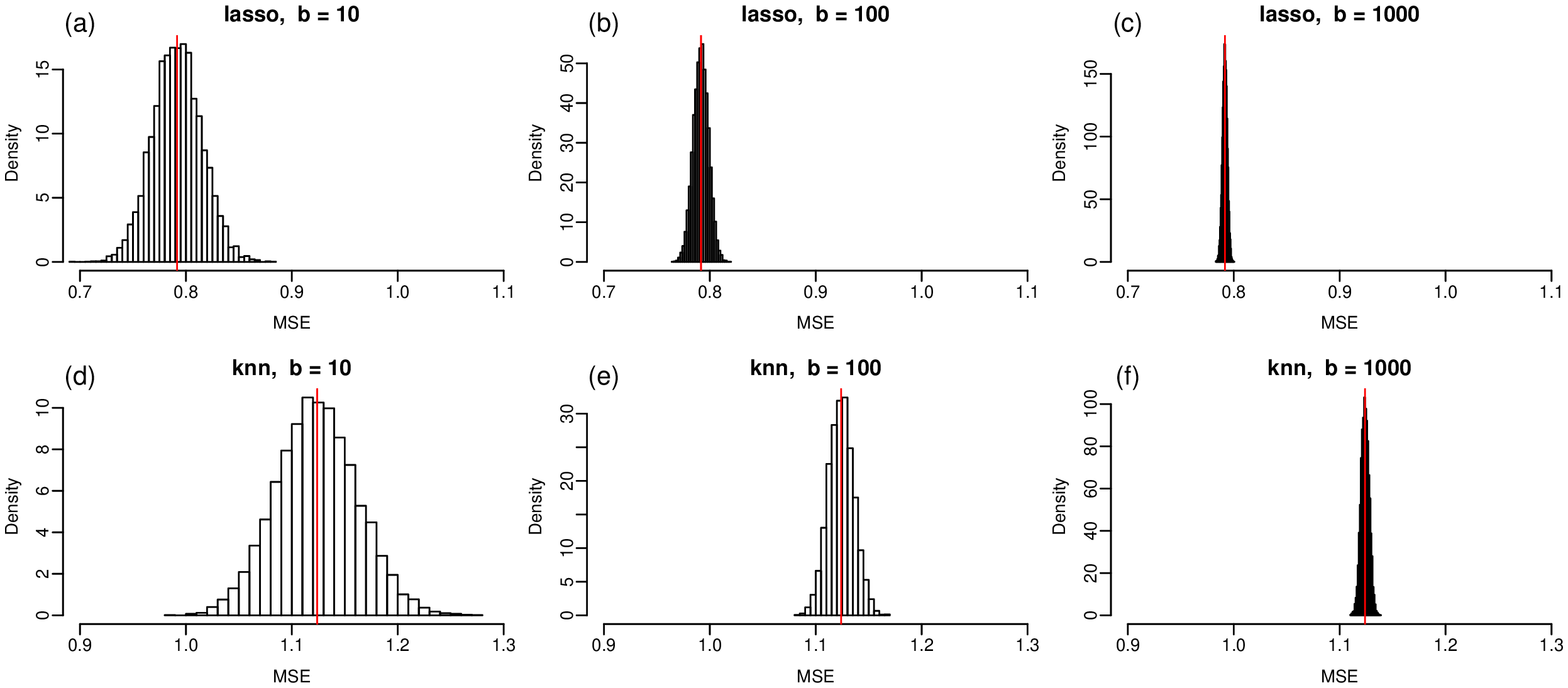}
\caption{Distributions of the LadderBoot released scores for a lasso and knn regression predictions. Results based on simulated data.}
\label{fig:lassovsknn}
\end{center}
\end{supplefig}

To further illustrate this point we show in Figure S\ref{fig:signalvsvariability}a boxplots of the distributions of the LadderBoot scores generated from eight increasing better quality predictions (with MSE scores varying from 1.64 to 0.15, and corresponding Pearson correlations varying from 0.18 to 0.92) for $b$ equal to 10, 100, and 1000. (In this example, the predictions were generated in an artificial fashion by simply adding decreasing amounts of gaussian noise to the true label data.) Figure S\ref{fig:signalvsvariability}b reports the standard deviations of the distributions. The results show a clear monotonic decrease in the variability of the LadderBoot released scores as a function of increasingly better predictions.

In both Figures S\ref{fig:lassovsknn} and S\ref{fig:signalvsvariability}, the distributions were generated from 10,000 replications of the following process: (i) bootstrap the prediction (and the corresponding true label) data $b$ times; (ii) compute the MSE score on each of the $b$ bootstrapped data sets; and (iii) compute the average MSE across the $b$ bootstrap scores.

\begin{supplefig}[!h]
\begin{center}
\includegraphics[angle=270, scale = 0.49, clip]{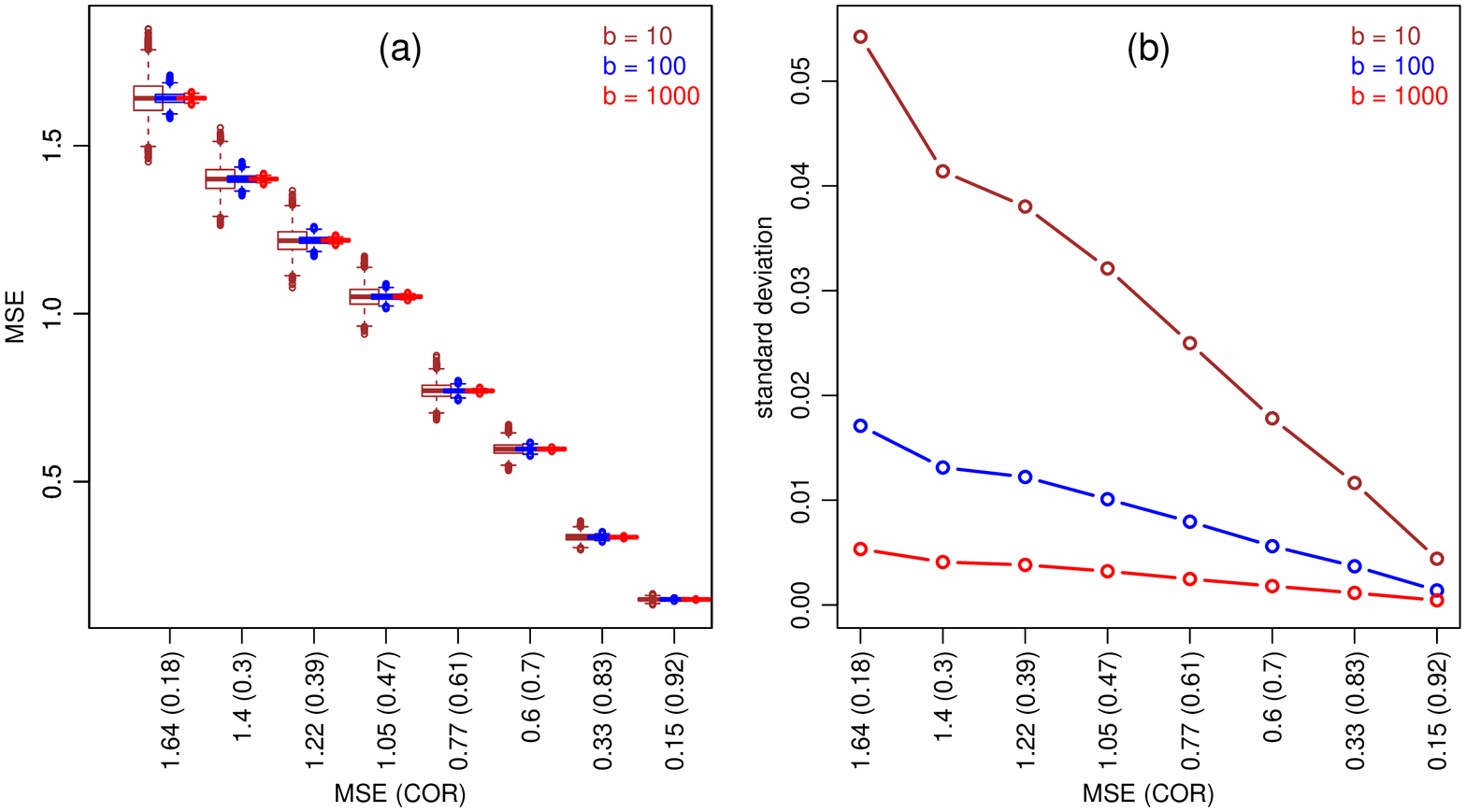}
\caption{Monotonic decrease in the variability of the LadderBoot released scores as a function of increasingly better predictions. The x-axis show the MSE scores (and respective Pearson correlation in parenthesis) for eight predictions of increasingly better quality.}
\label{fig:signalvsvariability}
\end{center}
\end{supplefig}

\end{document}